\documentclass[journal,onecolumn]{IEEEtran}

\ifCLASSINFOpdf
\else
   \usepackage[dvips]{graphicx}
\fi
\usepackage{url}

\usepackage{graphicx}

\usepackage{booktabs}
\usepackage{amsmath}  
\usepackage{amssymb}
\usepackage{amsxtra}      
\usepackage{multirow}
\usepackage{mathrsfs}
\usepackage{xcolor}
\usepackage{tabularx}
\usepackage{mathtools}
\usepackage{amsthm}
\usepackage{algpseudocode}
\usepackage{physics}
\usepackage{ragged2e}
\usepackage{bm}
\usepackage{multirow}
\usepackage[lined,ruled,linesnumbered]{algorithm2e}

\usepackage{footnote}
\makesavenoteenv{algorithm}
\usepackage{multirow}

\newtheoremstyle{mystyle}
{}
{}
{\itshape}
{}
{\bfseries}
{.}
{ }
{}
\theoremstyle{mystyle}

\begin{document}

\title{On Recursive State Estimation for Linear State-Space Models Having Quantized Output Data}

\author{Angel L. Cedeño, Ricardo Albornoz, Boris I. Godoy, Rodrigo Carvajal and Juan C. Agüero,
\thanks{This research was supported in part by the Chilean National Agency for Research and Development (ANID) Scholarship Program/doctorado nacional/2020-21202410, in part by the PIIC program of DGP at Universidad Técnica Federico Santa María No 062/2018 and 035/2021, and in part by the grants ANID-Fondecyt 1181158 and 11160375, ANID-Basal Project FB0008 (AC3E).}
\thanks{Angel L. Cedeño, Ricardo Albornoz and Juan C. Agüero are with the Electronics Engineering Department, Universidad Técnica Federico Santa María, Valparaíso 2390123, Chile, and also with the Advanced Center for Electrical and Electronic Engineering, AC3E, Valparaíso 2390123, Chile (e-mail: angel.cedeno@sansano.usm.cl; ricardo.albornoz13@sansano.usm.cl; jcarlos.aguero@gmail.com).}
\thanks{Boris I. Godoy are with the Department of Mechanical Engineering, Boston University, Boston, MA 02215 USA. (e-mail: bgodoy@bu.edu).}
\thanks{Rodrigo Carvajal are with the Escuela de Ingeniería Eléctrica, Pontificia Universidad Católica de Valparaíso, Brasil 2147, Valparaíso, Chile (e-mail: rodrigo.carvajal@pucv.cl).}
}

\maketitle

\begin{abstract}
	In this paper, we study the problem of estimating the state of a dynamic state-space system where the output is subject to quantization. We compare some classical approaches and a new development in the literature to obtain the filtering and smoothing distributions of the state conditioned to quantized data. The classical approaches include the Extended Kalman filter/smoother in which we consider an approximation of the quantizer non-linearity based on the arctan function, the quantized Kalman filter/smoother, the Unscented Kalman filter/smoother, and the Sequential Monte Carlo sampling method also called particle filter/smoother. We consider a new approach based on the Gaussian sum filter/smoother where the probability mass function of the quantized data given the state is modeled as an integral equation and approximated using Gauss-Legendre quadrature.  The Particle filter is addressed considering some resampling methods used to deal with the degeneracy problem. Also, the sample impoverishment caused by the resampling method is addressed by introducing diversity in the samples set using the Markov Chain Monte Carlo method. In this paper, we discuss the implementation of the aforementioned algorithms and the Particle filter/smoother implementation is studied by using different resampling methods combined with two Markov Chain algorithms. A numerical simulation is presented to analyze the accuracy of the estimation and the computational cost. 
\end{abstract}


\IEEEpeerreviewmaketitle

\section{Introduction}

\IEEEPARstart{I}{n} the last two decades, there has been a growing number of applications for sensors, networks, and sensor networks, where common problems include dealing with loss of information in signals measured with low-resolution sensors, or storing and/or transmitting a reduced representation of signals to minimize the resource consumption in a communication channel \cite{gersho2012}. This kind of problem encompasses a nonlinear process called \textit{Quantization}, which divides the input signal space into a finite or infinite (countable) number of subspaces and represents each of them by a single output value \cite{widrow2008quantization}. Applications with quantized data include networked control \cite{Li2011,Zhang2020}, fault detection \cite{Li2011,Zhang2021,Noshad2019,Huang2021}, cyber-physical systems \cite{Liu2020,Ding2021}, multitarget tracking \cite{Wang2017}, and system identification  \cite{curry1970estimation,gustafsson2009statistical,wang2010system,marelli2010scenario}, to mention a few. In these applications, a key part of the development process is the ability to estimate the state of a dynamical system conditioned to the available quantized observations. For instance, \cite{Rana2015} deals with the problem of state estimation and control of a microgrid incorporating multiple distributed energy resources, where the state estimation and control is based on uniform quantized observations that are transmitted through a wireless channel. 

In general, for nonlinear and non-Gaussian systems, it is impossible to obtain closed-form expressions of the filtering and smoothing probability density functions (PDFs) and in turn a state estimators. However, for linear and Gaussian systems, the optimal solution is given by the Kalman Filter (KF) and the Rauch-Tung-Striebel smoother (KS), respectively \cite{sarkka2013bayesian}. Many sub-optimal methods have been developed in order to obtain an approximation of the desired PDFs and state estimation, see e.g. \cite{anderson2007optimal,Julier1997,arasaratnam2007discrete}. The Extended Kalman filter (EKF) was studied in \cite{Sviestins2001} to deal with quantized data. This approach is difficult to implement since the quantizer is a non-differentiable non-linear function and requires the computation of a Jacobian matrix. The authors in \cite{Sviestins2001} proposed to approximate the quantizer by using a smooth function to compute approximately the Jacobian matrix. However, due to the highly non-linearity of the quantizer, the EKF/EKS approach produces non-accuracy estimates of the system state. The Kalman filter (KF) was modified to include the quantization effect in the computation of the filtering state, which has been referred to as quantized Kalman filter (QKF) \cite{Gomez2020,leong2013}. The Unscented Kalman filter (UKF) was applied by \cite{zhou2009} to deal with quantized innovation systems in a wireless sensor network environment. The UKF is based on the Unscented transformation \cite{Julier1997} that represents the mean and covariance of a Gaussian random variable through a reduced number of points. Then, these points are propagated through the nonlinear function to accurately capture the mean and covariance of the propagated Gaussian random variable. The advantage of the UKF is high estimation accuracy and convergence rate, and the simple implementation compared with EKF, avoiding the computation of the Jacobian matrix required in the EKF method. 

One of the most used methods in nonlinear filtering is the Sequential Monte Carlo sampling approach called particle filter (PF)\cite{Gordon1993}. PF uses a set of weights and samples to represent approximately the filtering and smoothing PDFs. The advantages of PF/PS are the facility of implementation and the ability to deal with highly nonlinear and non-Gaussian systems. However, particle filter suffers from some drawbacks. One of them is the degeneracy problem, which means that in every iteration of the particle filter algorithm most weights are going to zero \cite{Doucet2000}. To solve this problem, \cite{Gordon1993} proposed an approach called resampling, in which the heavily weighted particles are replicated sometimes, and the rest of the particles are unused. A number of resampling schemes have been developed in the literature such as multinomial, stratified, systematic, residual, branch-kill, rounding-copy, Metropolis, and local selection resampling methods, to mention a few \cite{Li2015b}. These resampling methods have some advantages such as unbiasedness and parallelism capacity. In addition, the performance of the particle filter depends on the used resampling method \cite{Douc2005}. Unfortunately, the resampling process produces a loss of diversity in the particle set since the particles with high weights are replicated. This problem is called sample impoverishment \cite{Doucet2000}, and to mitigate it, a MCMC move is usually introduced after the resampling step in order to provide diversity to the samples so that the new particle set is still distributed according to the posterior PDF \cite{Bi2015}. There are mainly two MCMC methods that we can use to deal with the impoverishment problem: the Gibbs and the Metropolis-Hasting (MH) sampling. Here we have used the MH algorithm and a special MH algorithm called Random-Walk Metropolis (RWM). 

On the other hand, a new algorithm to deal with quantized output data is proposed in \cite{Cedeno2021,Cedeno2021b}. In these works, the authors defined the probability mass function of the quantized data conditioned to the system state as an integral equation depending on the quantizer regions. This integral equation is solved utilizing Gauss-Legendre quadrature, which yields a model with Gaussian sum structure. This model is used to develop Gaussian sum filter/smoother to obtain closed-form filtering and smoothing distributions for systems with quantized data.

In this paper, we discuss the implementation of EKF/EKS,  QKF/QKS, UKF/UKS, GSF/GSS and PF/PS algorithms to deal with the problem of quantized observations. The PF/PS algorithm is studied by combining the resampling methods: systematic (SYS), multinomial (ML), Metropolis (MT), and local selection (LS) \cite{Li2015b}, with the MCMC algorithms: MH and RWM \cite{Zhai2004,Sherlock2015}. The analysis is carried out in terms of the accuracy of the state estimation and the computational cost. The organization of this paper is as follows: In Section \ref{sec:problem_def} the problem of interest is defined. In Section \ref{sec:recursive_fil_smoot} we present a brief review of the filtering and smoothing methods. In Section \ref{sec:numerical_experiments} we present a numerical example. Finally, the concluding remarks are given in section \ref{sec:conclusions}.

\section{Statement of the Problem}\label{sec:problem_def}

\begin{figure}[!hb] 
	\centerline{
	\includegraphics[width=0.5\linewidth]{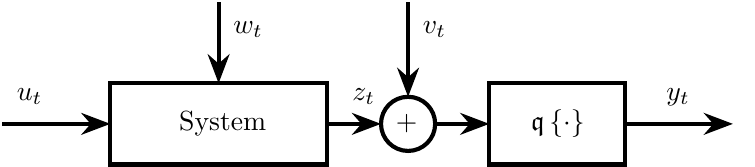}}
	\caption{State-space model with quantized output.}  
	\label{fig:diagram_ssq}                                                               
\end{figure}

This paper consider the filtering and smoothing problem for the following discrete-time time-invariant state-space system with quantized output (See Figure. \ref{fig:diagram_ssq}):
\begin{align}
	\mathbf{x}_{t+1}&=\mathbf{A}\mathbf{x}_{t}+\mathbf{B}\mathbf{u}_{t}+\mathbf{w}_{t}, \label{eqn:general_system_state}\\
	z_{t}&=\mathbf{C}\mathbf{x}_{t}+\mathbf{D}\mathbf{u}_{t}+v_{t}, \label{eqn:general_system_output}\\
	y_{t}&= \mathfrak{q} \left\lbrace z_{t} \right\rbrace \label{eqn:general_system_qoutput},
\end{align}
where $\mathbf{x}_{t} \in \mathbb{R}^{n}$ is the state vector, $z_{t} \in \mathbb{R}$ is the non-quantized output, $y_{t} \in \mathbb{R}$ is the quantized output, and $\mathbf{u}_{t} \in \mathbb{R}^{m}$ is the input of the system. The matrix $\mathbf{A} \in \mathbb{R}^{n\times n}$, $\mathbf{B} \in \mathbb{R}^{n\times m}$, $\mathbf{C} \in \mathbb{R}^{1\times n}$ and $\mathbf{D} \in \mathbb{R}^{1\times m}$. The nonlinear map $\mathfrak{q}\left\lbrace \cdot\right\rbrace$ is the quantizer. The state noise $\mathbf{w}_{t} \in \mathbb{R}^{n}$ and the output noise $v_{t} \in \mathbb{R}$ are zero-mean white Gaussian noises with covariance matrix $\mathbf{Q}$ and $R$, respectively. Due to the random components (i.e., the noises $\mathbf{w}_{t}$ and $v_{t}$) in \eqref{eqn:general_system_state} and \eqref{eqn:general_system_output}, the state-space model can be described using the state transition PDF $p(\mathbf{x}_{t+1}|\mathbf{x}_t)\sim\mathcal{N}(\mathbf{x}_{t+1};\mathbf{A}\mathbf{x}_t+\mathbf{B}\mathbf{u}_t,\mathbf{Q})$ and the non-quantized output PDF $p(z_{t}|\mathbf{x}_t)\sim\mathcal{N}(z_t;\mathbf{C}\mathbf{x}_t+\mathbf{D}\mathbf{u}_t,R)$ with $\mathbf{x}_1\sim\mathcal{N}(\mathbf{x}_1;\bm{\mu}_1,\mathbf{P}_1)$, where $\mathcal{N}(\mathbf{x};\bm{\mu},\mathbf{P})$ represents a PDF corresponding to a Gaussian distribution with mean $\bm{\mu}$ and covariance matrix $\mathbf{P}$ of the variable $\mathbf{x}$. The initial condition $\mathbf{x}_1$, the model noise $\mathbf{w}_t$, and the measurement noise $v_t$ are statistically independent random variables. \\

On the other hand, we consider the quantizer  $\mathfrak{q}\left\lbrace \cdot\right\rbrace: \mathbb{R}\rightarrow \Psi$ where $\mathbb{R}$ is the real line and $\Psi\subset \mathbb{R}$ is the output set, then $\mathfrak{q}\left\lbrace \cdot\right\rbrace$ is given by \cite{gersho2012}:
\begin{align}
	y_t=\mathfrak{q}\left\lbrace z_t\right\rbrace=\left\lbrace \psi_k\in\Psi,\mathcal{R}_k\subset\mathbb{R}; k\in \mathcal{K} \right\rbrace \label{eqn:general_quantizer},
\end{align}
where $\psi_k$ is the $k$th output value in the output set $\Psi$, $\mathcal{R}_k$ is the $k$th interval mapped to the value $\psi_k$, and the indices set $\mathcal{K}$ defines the number of quantization levels of the output set $\Psi$. Here we consider two types of quantizers: (i) an infinite-level quantizer (ILQ), in which the output of the quantizer has infinite (countable) levels of quantization with
\begin{align}
	\mathcal{K}=\left\lbrace\dots, 1,2, \dots, L,\dots \right\rbrace, \label{eqn:index_infinite}
\end{align}
where $\mathcal{R}_k=\left\lbrace z_t: q_{k-1} \leq z_t < q_k\right\rbrace$ are disjoint intervals, and each $\psi_k$ is the value that the quantizer takes in the region $\mathcal{R}_k$, and (ii) a finite-level quantizer (FLQ), in which the output of the quantizer is limited to minimum and maximum values (saturated quantizer) similar to \eqref{eqn:general_quantizer} with
\begin{align}
	\mathcal{K}=\left\lbrace 1,2, \dots, L-1,L \right\rbrace. \label{eqn:index_finite}
\end{align}
Notice that the FLQ quantizer is comprised with finite and semi-infinite intervals given by $\mathcal{R}_{1}=\left\lbrace z_t: z_t < q_1\right\rbrace $, $\mathcal{R}_{L}=\left\lbrace z_t: q_{L-1} \leq z_t \right\rbrace $, and $\mathcal{R}_k=\left\lbrace z_t: q_{k-1} \leq z_t < q_k\right\rbrace$, with $k=2\dots,L-1$.

Thus, the problem of interest can be defined as follows: Given the available data  $\mathbf{u}_{1:N}=\left\lbrace \mathbf{u}_1,\mathbf{u}_2,\dots,\mathbf{u}_N\right\rbrace$ and $y_{1:N}=\left\lbrace y_1,y_2,\dots,y_N\right\rbrace$, where $N$ is the data length, obtain the filtering and smoothing PDFs of the state given the quantized measurements, $p(\mathbf{x}_t | y_{1:t})$ and $p(\mathbf{x}_t|y_{1:N})$, respectively, the state estimators:
\begin{align}
	\hat{\mathbf{x}}_{t|t}&=\mathbb{E}\left\lbrace \mathbf{x}_t|y_{1:t}\right\rbrace=\int\mathbf{x}_tp(\mathbf{x}_t|y_{1:t})d\mathbf{x}_t, \label{eqn:filtered_state} \\
	\hat{\mathbf{x}}_{t|N}&=\mathbb{E}\left\lbrace \mathbf{x}_t|y_{1:N}\right\rbrace=\int\mathbf{x}_tp(\mathbf{x}_t|y_{1:N})d\mathbf{x}_t, \label{eqn:smoothed_state}
\end{align}
and the corresponding covariance matrices of the estimation error:
\begin{align}
	\Sigma_{t|t}&=\mathbb{E}\left\lbrace (\mathbf{x}_t-\hat{\mathbf{x}}_{t|t})(\mathbf{x}_t-\hat{\mathbf{x}}_{t|t})^T|y_{1:t}\right\rbrace=\int (\mathbf{x}_t-\hat{\mathbf{x}}_{t|t})(\mathbf{x}_t-\hat{\mathbf{x}}_{t|t})^Tp(\mathbf{x}_t|y_{1:t})d\mathbf{x}_t, \label{eqn:filtered_cov_state}\\
	\Sigma_{t|N}&=\mathbb{E}\left\lbrace (\mathbf{x}_t-\hat{\mathbf{x}}_{t|N})(\mathbf{x}_t-\hat{\mathbf{x}}_{t|N})^T|y_{1:N}\right\rbrace=\int (\mathbf{x}_t-\hat{\mathbf{x}}_{t|t})(\mathbf{x}_t-\hat{\mathbf{x}}_{t|t})^Tp(\mathbf{x}_t|y_{1:N})d\mathbf{x}_t, \label{eqn:smoothed_cov_state}
\end{align}
where $t\leq N$ and $\mathbb{E}\left\lbrace \mathbf{x}|y\right\rbrace$ denotes the conditional expectation of $\mathbf{x}$ given $y$.
\section{Recursive Filtering and Smoothing Methods for quantized output data}\label{sec:recursive_fil_smoot}
\subsection{Bayesian filtering and smoothing}
From Bayesian approach, the filtering distribution admits the following recursion, see e.g. \cite{sarkka2013bayesian}:
\begin{align}
	p(\mathbf{x}_{t}|y_{1:t-1})&=\int p(\mathbf{x}_{t}|\mathbf{x}_{t-1})p(\mathbf{x}_{t-1}|y_{1:t-1})d\mathbf{x}_{t-1}, \label{eqn:bayesian_filtering_time}\\
	p(\mathbf{x}_t|y_{1:t})&=\dfrac{p(y_t|\mathbf{x}_t)p(\mathbf{x}_t|y_{1:t-1})}{p(y_t|y_{1:t-1})}, \label{eqn:bayesian_filtering_measurement}
\end{align}
where $p(\mathbf{x}_t|y_{1:t})$ and $p(\mathbf{x}_{t}|y_{1:t-1})$ are the measurement- and time-update equations. The PDF $p(\mathbf{x}_{t}|\mathbf{x}_{t-1})$ is directly obtained from the model in \eqref{eqn:general_system_state}, and $p(y_t|y_{1:t-1})$ is a normalization constant. On the other hand, the  Bayesian smoothing equation, see e.g. \cite{sarkka2013bayesian}, is defined by:
\begin{equation}\label{eqn:forward_backward_smoothing}
	p(\mathbf{x}_{t}|y_{1:N})=p(\mathbf{x}_{t}|y_{1:t}) \! \int \frac{p(\mathbf{x}_{t+1}|y_{1:N})p(\mathbf{x}_{t+1}|\mathbf{x}_{t})}{p(\mathbf{x}_{t+1}|y_{1:t})}d\mathbf{x}_{t+1}.
\end{equation}
Notice that to obtain $p(\mathbf{x}_{t}|y_{1:t})$ in \eqref{eqn:bayesian_filtering_measurement} we need the probability function $p(y_t|\mathbf{x}_t)$. Since $y_t$ is a discrete random variable, the probabilistic model of $p(y_t|\mathbf{x}_t)$ is a PMF. Then, the measurement-update equation in \eqref{eqn:bayesian_filtering_measurement} combines both PDFs and a PMF. Here we use \textit{Generalized Probability Density Functions}, see e.g. \cite{degroot2005}, which comprises both discrete and absolutely continuous distributions. In \cite{Cedeno2021b}, an integral equation for $p(y_t|\mathbf{x}_t)$ is defined in order to solve the filtering recursion as follows
\begin{equation}\label{eqn:integral_thm_pytxt}
	p(y_t|\mathbf{x}_t) = \int_{a_t}^{b_t}\mathcal{N}\left(v_t;0,R \right) dv_t,
\end{equation}
where $a_t$ and $b_t$ are functions of the boundary values of each region of the quantizers defined in Table \ref{tab:my-table}.
\begin{center}
	\begin{table}[!ht]
		\normalsize
		\caption{Integral limits  of equation \eqref{eqn:integral_thm_pytxt}.}
		\label{tab:my-table}
		\setlength{\tabcolsep}{10pt}
		\setlength{\extrarowheight}{3pt}
		\centerline{
		\begin{tabular}{|c|c|c|c|}
			\hline 
			\multirow{4}{*}{FLQ}
			&$y_t$       & $a_t$ & $b_t$ \\ \cline{2-4} 
			&$\psi_{1}$& $-\infty$  & $q_1-\mathbf{C}\mathbf{x}_t-\mathbf{D}\mathbf{u}_t$   \\ \cline{2-4} 
			&$\begin{matrix}
				\psi_k,\\k=2,\dots,L-1
			\end{matrix}$ & $q_{k-1}-\mathbf{C}\mathbf{x}_t-\mathbf{D}\mathbf{u}_t$  & $q_k-\mathbf{C}\mathbf{x}_t-\mathbf{D}\mathbf{u}_t$  \\ \cline{2-4} 
			&$\psi_{L}$ & $q_{L-1}-\mathbf{C}\mathbf{x}_t-\mathbf{D}\mathbf{u}_t$  &  $\infty$  \\ \hline 
			ILQ
			&$\begin{matrix}
				\psi_k,\\ k=\dots,1,\dots,L,\dots
			\end{matrix}$ & $q_{k-1}-\mathbf{C}\mathbf{x}_t-\mathbf{D}\mathbf{u}_t$   &  $q_k-\mathbf{C}\mathbf{x}_t-\mathbf{D}\mathbf{u}_t$  \\ \hline 
		\end{tabular}}
	\end{table}
\end{center}
Notice that $y_t|\mathbf{x}_t$ in equation \eqref{eqn:integral_thm_pytxt} is a non-Gaussian random variable. This leads to obtain non-Gaussian measurement- and time-update distributions. However, the EKF, QKF, and UKF filters are developed under the assumption that the measurement- and time-update distributions are Gaussian, which yields a loss of accuracy in the estimation. On the other hand, the equation \eqref{eqn:integral_thm_pytxt} is used in GSF/GSS and PF/PS, where the Gaussian assumption is not considered. 
\subsection{Extended state-space system}
To implement some filtering and smoothing algorithms such as the EKF/EKS and the UKF/UKS, the state-space model in \eqref{eqn:general_system_state}-\eqref{eqn:general_system_qoutput} is rewritten in a extended form as follows
\begin{align}
	\mathbf{x}_{t+1}^{\textrm{e}}&=\mathcal{A}\mathbf{x}_t^{\textrm{e}}+\mathcal{B}\mathbf{u}_t^{\textrm{e}}+\mathbf{w}_t^{\textrm{e}}, \label{eqn:exteded_system_state}\\
	y_{t}&= \mathfrak{q} \left\lbrace \mathcal{C}\mathbf{x}_{t}^{\textrm{e}} \right\rbrace +\xi_t,\label{eqn:exteded_system_qoutput}
\end{align}
where the extended system matrices are given by 
\begin{equation}
	\mathcal{A}=\left[\begin{matrix}\mathbf{A}&0\\\mathbf{A}\mathbf{C}&0\end{matrix}\right], \qquad \mathcal{B}=\left[\begin{matrix}\mathbf{B}&0\\\mathbf{C}\mathbf{B}&\mathbf{D}	\end{matrix}\right], \qquad \mathcal{C}=[\begin{matrix}0&1\end{matrix}],
\end{equation}
where $\mathbf{x}_{t}^{\textrm{e}}=[ \begin{matrix} \mathbf{x}_{t}^T& z_{t}\end{matrix}]^T$ is the extended state, $\mathbf{u}_t^{\textrm{e}}=[\begin{matrix} \mathbf{u}_t^T&\mathbf{u}_{t+1}^T\end{matrix}]$ is the extended input with $\mathbf{u}_{N+1}=0$, and the initial condition $\mathbf{x}_1^{\textrm{e}}\sim\mathcal{N}(\mathbf{x}_1^{\textrm{e}};\bm{\mu}_1^{\textrm{e}},\mathbf{P}_1^{\textrm{e}})$ where $\bm{\mu}_1^{\textrm{e}}=[\bm{\mu}_1^T~~0]^T$ and $\mathbf{P}_1^{\textrm{e}}=\textrm{diag}\left\lbrace  \mathbf{P}_1,1\right\rbrace $. The noise $\mathbf{w}_t^{\textrm{e}}=[\begin{matrix} \mathbf{w}_t^T& (\mathbf{C}\mathbf{w}_t+v_{t+1})^T\end{matrix}]^T$ satisfies $\mathbf{w}_t^{\textrm{e}} \sim \mathcal{N}\left(\mathbf{w}_t^{\textrm{e}};0,\mathcal{Q}\right)$ with
\begin{equation}
	\mathcal{Q}=\left[\begin{matrix} \mathbf{Q}&\mathbf{Q}\mathbf{C}^T\\\mathbf{C}\mathbf{Q}^T&\mathbf{C}\mathbf{Q}\mathbf{C}^T+R\end{matrix}\right].
\end{equation}
The noise $\xi_t$ is added in order to implement the EKF/EKS and the UKF/UKS. We assume that  $\xi_t\sim\mathcal{N}\left(\xi_t;0,\varepsilon\right)$, where the variance $\varepsilon$ is small. 
\subsection{Extended Kalman Filtering and Smoothing}
\begin{figure}[!ht]
	\centerline{
		\includegraphics[width=1\linewidth]{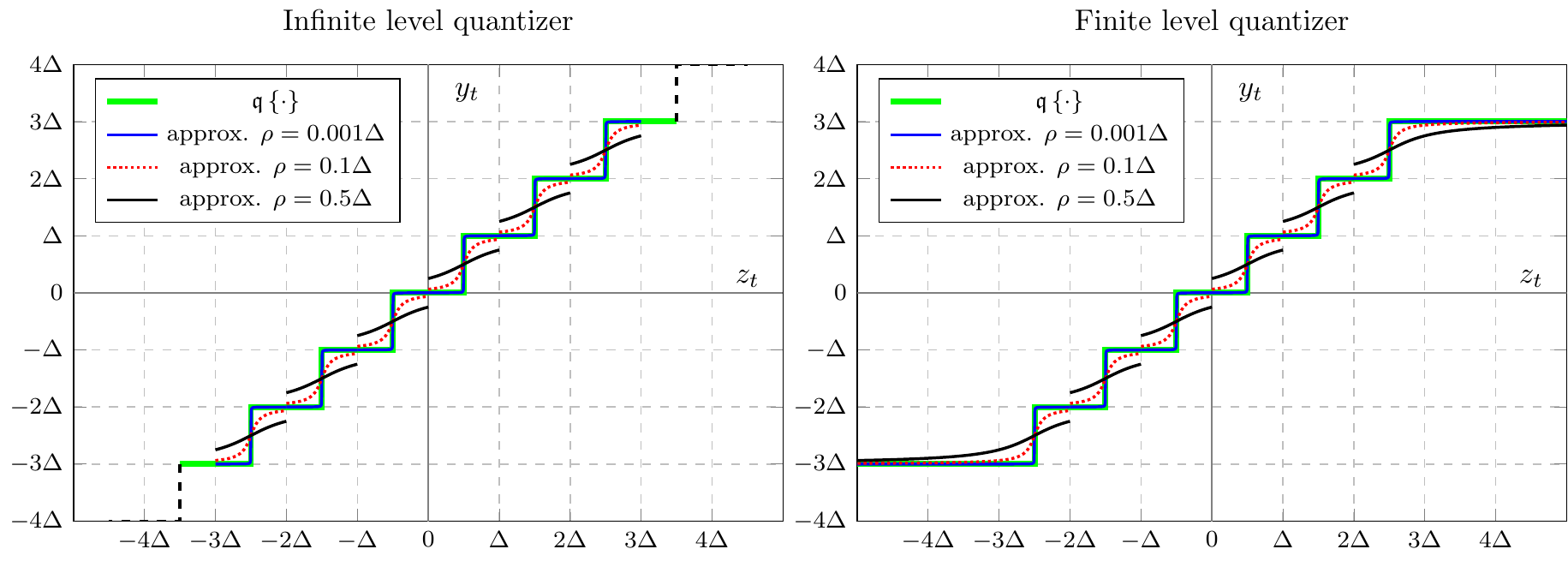}}
	\caption{Quantizer approximation by using the arctan function.}  
	\label{fig:quantizer_approximation}                                                               
\end{figure}

The extended Kalman filter \cite{Gelb1974,Grewal2014} is an extension of the standard Kalman filter to nonlinear state-space systems with process and measurement Gaussian noises. The idea of EKF is to build a linear approximation around a state estimation by using a Taylor series expansion. The EKF is not directly applied to the problem of interest in this paper since the quantizer is a non-differentiable nonlinear function. In \cite{Sviestins2001}, it was suggested that it is possible to compute the Kalman gain using a smooth \textit{arctan}-based approximation of the quantizer. Following the idea in \cite{Sviestins2001} and the representation of the arctan function found in \cite{Traore2014}, the following approximation of the quantizer is proposed:
\begin{equation}\label{eqn:quantizer_approximation}
	\mathfrak{q} \left\lbrace z_{t} \right\rbrace \approx h(z_t)=
	\left\lbrace 
	\begin{matrix}
		\vdots&&  && \vdots \\
		(\Delta/\pi)\atan\left\lbrace (z_t-2.5\Delta)/\rho\right\rbrace+2.5\Delta && \textrm{if} && 2\Delta\leq z_t<3\Delta,\\
		(\Delta/\pi)\atan\left\lbrace (z_t-1.5\Delta)/\rho\right\rbrace+1.5\Delta && \textrm{if} && \Delta\leq z_t<2\Delta,\\
		(\Delta/\pi)\atan\left\lbrace (z_t-0.5\Delta)/\rho\right\rbrace+0.5\Delta && \textrm{if} && 0\leq z_t<\Delta,\\
		(\Delta/\pi)\atan\left\lbrace (z_t+0.5\Delta)/\rho\right\rbrace-0.5\Delta && \textrm{if} && -\Delta\leq z_t<0,\\
		(\Delta/\pi)\atan\left\lbrace (z_t+1.5\Delta)/\rho\right\rbrace-1.5\Delta && \textrm{if} && -2\Delta\leq z_t<-\Delta,\\
		(\Delta/\pi)\atan\left\lbrace (z_t+2.5\Delta)/\rho\right\rbrace-2.5\Delta && \textrm{if} && -3\Delta\leq z_t<-2\Delta,\\
		\vdots&&  && \vdots \\
	\end{matrix}
	\right.
\end{equation}
where $\rho$ is a user parameter that defines how the approximation fits the quantizer function in the switch point, as is shown in Figure \ref{fig:quantizer_approximation}. 
\begin{figure}[!ht]
	\centerline{
		\includegraphics[width=1\linewidth]{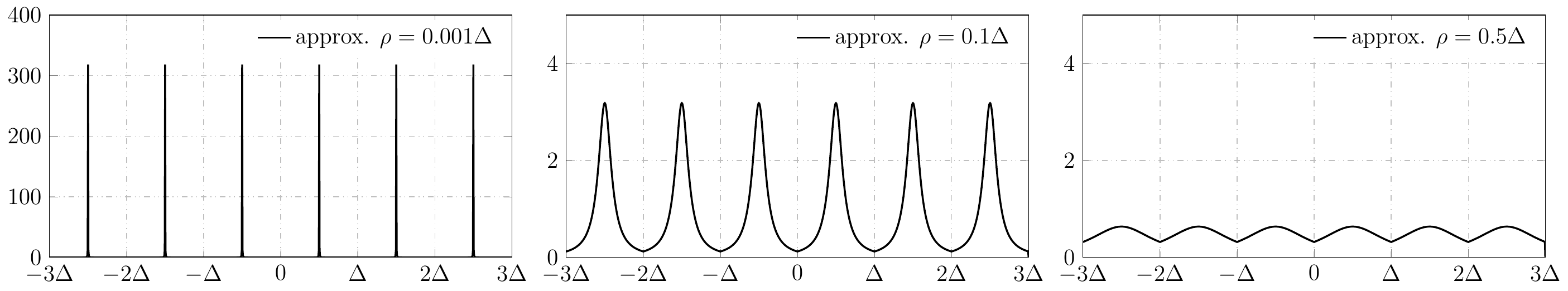}}
	\caption{Jacobian matrix of the quantizer approximation.}  
	\label{fig:quantizer_approximation_jac}                                                               
\end{figure}
On the other hand, using the smooth approximation of the quantizer it is possible to rewrite the nonlinear system as a type of linear time-varying system as follows: 
\begin{align}
\mathbf{x}_{t+1}^{\textrm{e}}&=\mathcal{A}\mathbf{x}_t^{\textrm{e}}+\mathcal{B}\mathbf{u}_t^{\textrm{e}}+\mathbf{w}_t^{\textrm{e}}, \label{eqn:ekf_approx_system_state}\\
	y_{t}&=\mathbf{H}_t\mathbf{x}_{t}^{\textrm{e}}+\mathbf{F}_t+\xi_{t}, \label{eqn:ekf_approx_system_output}
\end{align}
where $\mathbf{H}_t$ is the Jacobian matrix of $h(\mathcal{C}\mathbf{x}_{t}^{\textrm{e}})$ with respect to $\mathbf{x}_{t}^{\textrm{e}}$ and evaluated at $\hat{\mathbf{x}}_{t|t-1}^{\textrm{e}}$, $\mathbf{F}_t=h(\mathcal{C}\hat{\mathbf{x}}_{t|t-1}^{\textrm{e}}) -\mathbf{H}_t\hat{\mathbf{x}}_{t|t-1}^{\textrm{e}}$. Then, the equations of the EKF are summarized in Algorithm \ref{alg:extended_kalman_filter}. One of the difficulties in applying the EKF to deal with quantized data is the computation on the Jacobian matrix $\mathbf{H}_t$. Despite the approximation of the quantizer, the Jacobian is nearly zero for all values of  $\hat{\mathbf{x}}_{t|t-1}^{\textrm{e}}$, except for the exact switch points as shown in Figure \ref{fig:quantizer_approximation_jac} (left) where $\rho=0.001\Delta$. Additionally, Figure \ref{fig:quantizer_approximation_jac} (center and right) shows how the quantizer and Jacobian approximations worsen as $\rho$ increases, which reduces the accuracy of the estimations.
\begin{algorithm}
	\label{alg:extended_kalman_filter} 
	\SetKw{Inicio}{Input:}
	\Inicio{}  The distribution of the initial condition $p(\mathbf{x}_1^{\textrm{e}})$, e.g. $\hat{\mathbf{x}}_{0|1}^{\textrm{e}}=\bm{\mu}_1^{\textrm{e}}$ and $\Sigma_{0|1}^{\textrm{e}}=\mathbf{P}_1^{\textrm{e}}$. \\
	\For{$t=1$ \textrm{\textbf{to}} $N$}{
		Compute the Kalman gain using: $\mathbf{K}_t=\Sigma_{t|t-1}^{\textrm{e}}\mathbf{H}_t^T\left(\varepsilon+\mathbf{H}_t\Sigma_{t|t-1}^{\textrm{e}}\mathbf{H}_t^T\right)^{-1}$.\\
		\textbf{Measurement Update:}\\
		Compute the filtering state $\hat{\mathbf{x}}_{t|t}^{\textrm{e}}$ according to $\hat{\mathbf{x}}_{t|t}^{\textrm{e}}= \hat{\mathbf{x}}_{t|t-1}^{\textrm{e}} + \mathbf{K}_t\left(y_t-h\left(\mathcal{C}\hat{\mathbf{x}}_{t|t-1}^{\textrm{e}}\right)\right)$.\\
		Compute the covariance matrix $\Sigma_{t|t}^{\textrm{e}}$ according to $\Sigma_{t|t}^{\textrm{e}}= (\mathbf{I}-\mathbf{K}_t\mathbf{H}_t)\Sigma_{t|t-1}^{\textrm{e}}$.\\
		\textbf{Time Update:}\\
		Compute the predicted state $\hat{\mathbf{x}}_{t+1|t}^{\textrm{e}}$ according to $\hat{\mathbf{x}}_{t+1|t}^{\textrm{e}} = \mathcal{A}\hat{\mathbf{x}}_{t|t}^{\textrm{e}}+\mathcal{B}\mathbf{u}_t^{\textrm{e}}$.\\
		Compute the covariance matrix $\Sigma_{t+1|t}^{\textrm{e}}$ according to $\Sigma_{t+1|t}^{\textrm{e}} = \mathcal{Q} + \mathcal{A}\Sigma_{t|t}^{\textrm{e}}\mathcal{A}^T$.
	}
	\SetKw{Fin}{Output:}
	\Fin{} The PDF $p(\mathbf{x}_t^{\textrm{e}}|y_{1:t})\sim \mathcal{N}\left(\mathbf{x}_t^{\textrm{e}}; \hat{\mathbf{x}}_{t|t}^{\textrm{e}},\Sigma_{t|t}^{\textrm{e}}\right) $ and the PDF $p(\mathbf{x}_{t+1}^{\textrm{e}}|y_{1:t})\sim \mathcal{N}\left(\mathbf{x}_{t+1}^{\textrm{e}}; \hat{\mathbf{x}}_{t+1|t}^{\textrm{e}},\Sigma_{t+1|t}^{\textrm{e}}\right)$ for $t=1,\dots,N$.
	\caption{Extended Kalman Filter} 
\end{algorithm}

\begin{algorithm}
	\label{alg:extended_kalman_smoother}
	\SetKw{Inicio}{Input:}
	\Inicio{}  The PDF $p(\mathbf{x}_t^{\textrm{e}}|y_{1:t})\sim \mathcal{N}\left(\mathbf{x}_t^{\textrm{e}}; \hat{\mathbf{x}}_{t|t}^{\textrm{e}},\Sigma_{t|t}^{\textrm{e}}\right) $ and the PDF $p(\mathbf{x}_{t+1}^{\textrm{e}}|y_{1:t})\sim \mathcal{N}\left(\mathbf{x}_{t+1}^{\textrm{e}}; \hat{\mathbf{x}}_{t+1|t}^{\textrm{e}},\Sigma_{t+1|t}^{\textrm{e}}\right)$ for $t=1,\dots,N$ computed in Algorithm \ref{alg:extended_kalman_filter}. \\
	\For{$t=N$ \textrm{\textbf{to}} $1$}{
		Compute the gain $\mathbf{G}_t=\Sigma_{t|t}^{\textrm{e}}\mathcal{A}_t^T\left(\Sigma_{t+1|t}^{\textrm{e}}\right)^{-1}$.\\
		Compute the smoothing state $\hat{\mathbf{x}}_{t|N}^{\textrm{e}}$ according to $\hat{\mathbf{x}}_{t|N}^{\textrm{e}}= \hat{\mathbf{x}}_{t|t}^{\textrm{e}} + \mathbf{G}_t\left( \hat{\mathbf{x}}_{t+1|T}^{\textrm{e}}-\hat{\mathbf{x}}_{t+1|t}^{\textrm{e}}\right)$.\\
		Compute the covariance matrix $\Sigma_{t|N}^{\textrm{e}}$ according to $\Sigma_{t|N}^{\textrm{e}}= \Sigma_{t|t}^{\textrm{e}} + \mathbf{G}_t\left( \Sigma_{t+1|T}^{\textrm{e}}-\Sigma_{t+1|t}^{\textrm{e}}\right)\mathbf{G}_t^{T}$.
	}
	\SetKw{Fin}{Output:}
	\Fin{} The PDF $p(\mathbf{x}_t^{\textrm{e}}|y_{1:N})\sim \mathcal{N}\left(\mathbf{x}_t^{\textrm{e}}; \hat{\mathbf{x}}_{t|N}^{\textrm{e}},\Sigma_{t|N}^{\textrm{e}}\right)$ for $t=1,\dots,N$.
	\caption{Extended Kalman Smoother} 
\end{algorithm}
\subsection{Unscented Kalman Filtering and Smoothing}
The unscented Kalman Filter \cite{Julier1997} is a deterministic-sampling-based approach that uses samples called sigma-points to propagate the mean and covariance of the system state (assumed to be a Gaussian random variable) through the nonlinear functions of the system. These propagated points accurately capture the mean and covariance of the posterior state to the 3rd order Taylor series expansion for any nonlinear function \cite{Wan2000}. The key idea of UKF is directly approximating the mean and covariance of the posterior distribution instead of approximating nonlinear functions \cite{Julier1997}. The unscented Kalman Filter is based on the unscented transformation of the random variable $x$ into the random variable $y$ as follows: $y=g(x)+v$ where $g(\cdot)$ is a nonlinear function, $x\sim\mathcal{N}\left(x;\mathbf{m},\Gamma\right)$, and $v\sim\mathcal{N}\left(v;0,\mathbf{P}\right)$, thus the sigma-points are defined by
\begin{align}
	\psi^{0}&=\mathbf{m},\\
	\psi^{\tau}&=\mathbf{m}+\sqrt{n+\lambda}\left[\sqrt{\mathbf{P}}\right]_{\tau}, \\
	\psi^{\tau+n}&=\mathbf{m}-\sqrt{n+\lambda}\left[\sqrt{\mathbf{P}}\right]_{\tau} ,
\end{align}
where $\tau=1,\dots,n$, the scaling parameter $\lambda=\alpha^2(n+\kappa)-n$, the parameters $\alpha$ and $\kappa$ determine the propagation of the sigma-points around the mean. The notation $\left[\mathcal{P}\right]_{\tau}$ refers to the $\tau$th column of the matrix $\mathcal{P}$ and $\sqrt{\mathcal{P}}$ represents the matrix square root of the matrix $\mathcal{P}$ such that $\sqrt{\mathcal{P}}\sqrt{\mathcal{P}}^{T}=\mathcal{P}$. The weights associated to the unscented transformation are the sets
\begin{align}
	\left\lbrace \omega^{0},\omega^{1},\dots,\omega^{2n}\right\rbrace &=\left\lbrace \lambda\zeta,0.5\zeta,\dots,0.5\zeta \right\rbrace, \\
	\left\lbrace \sigma^{0},\sigma^{1},\dots,\sigma^{2n}\right\rbrace&=\left\lbrace  \lambda\zeta+\varrho,0.5\zeta,\dots,0.5\zeta\right\rbrace ,
\end{align}
where $\zeta=(n+\lambda)^{-1}$, $\varrho=1-\alpha^2+\beta$. Here $\beta$ is an additional algorithm parameter that can be used for incorporating prior information on the (non-Gaussian) distribution of $x$. Then, the statistics of the transformed random variable are: the mean $\mu=\sum_{\tau=0}^{2n}\omega^{\tau}g(\psi^{\tau})$ and the covariance matrix $\Phi=\sum_{\tau=0}^{2n}\sigma^{\tau}\left[ g(\psi^{\tau})-\mu\right]\left[ g(\psi^{\tau})-\mu\right]^T+\mathbf{P}$. Additionally, the cross covariance matrix between $x$ and $y$ is given by $\Psi=\sum_{\tau=0}^{2n}\sigma^{\tau}\left[ \psi^{\tau}-\mathbf{m}\right]\left[ g(\psi^{\tau})-\mu\right]^T$.  The steps to implement the UKF are summarized in Algorithm \ref{alg:unscented_kalman_filter}. Notice that for the problem of interest in this paper, the process equation is a linear function. Thus, the UKS algorithm is similar to EKS but using the filtering and predictive distributions obtained with the UKF algorithm. 
\begin{algorithm}
	\label{alg:unscented_kalman_filter}
	\SetKw{Inicio}{Input:}
	\Inicio{}  The distribution of the initial condition $p(\mathbf{x}_1^{\textrm{e}})$, e.g. $\hat{\mathbf{x}}_{0|1}^{\textrm{e}}=\bm{\mu}_1^{\textrm{e}}$ and $\Sigma_{0|1}^{\textrm{e}}=\mathbf{P}_1^{\textrm{e}}$, the constant $\alpha$, $\kappa$, and $\beta$. \\
	\For{$t=1$ \textrm{\textbf{to}} $N$}{
		Compute and store the sigma-points $\psi_{t|t-1}^{\tau}$, the weigths $\omega_{t|t-1}^{\tau}$ and $\sigma_{t|t-1}^{\tau}$ by using $\hat{\mathbf{x}}_{t|t-1}^{\textrm{e}}$ and $\Sigma_{t|t-1}^{\textrm{e}}$ for $\tau=0,\dots,2n$. \\
		Propagate sigma-points using the measurement model $\mathcal{Y}_{t}^{\tau}= \mathfrak{q} \left\lbrace \mathcal{C}\psi_{t|t-1}^{\tau}\right\rbrace $ for $\tau=0,\dots,2n$.\\
		Compute the gain $\mathbf{K}_t=\Gamma_t\mathbf{S}_t^{-1}$, where
		\begin{align*}
			\bm{\nu}_t &= \sum_{\tau=1}^{2n} \omega_{t|t-1}^{\tau}\mathcal{Y}_{t}^{\tau}.\\
			\mathbf{S}_t &= \sum_{\tau=1}^{2n} \sigma_{t|t-1}^{\tau}\left(\mathcal{Y}_{t}^{\tau}- \bm{\nu}_t\right)\left(\mathcal{Y}_{t}^{\tau}- \bm{\nu}_t\right)^{T}+\varepsilon. \\
			\Gamma_t &= \sum_{\tau=1}^{2n} \sigma_{t|t-1}^{\tau}\left(\psi_{t|t-1}^{\tau}-\hat{\mathbf{x}}_{t|t-1}^{\textrm{e}}\right)\left(\mathcal{Y}_{t}^{\tau}- \bm{\nu}_t\right)^{T}.
		\end{align*}
		\textbf{Measurement Update:}\\
		Compute the filtering state $\hat{\mathbf{x}}_{t|t}^{\textrm{e}}$ according to $\hat{\mathbf{x}}_{t|t}^{\textrm{e}}= \hat{\mathbf{x}}_{t|t-1}^{\textrm{e}} + \mathbf{K}_t\left(y_t-\bm{\nu}_t\right)$.\\
		Compute the covariance matrix $\Sigma_{t|t}^{\textrm{e}}$ according to $\Sigma_{t|t}^{\textrm{e}}= \Sigma_{t|t-1}^{\textrm{e}}-\mathbf{K}_t\mathbf{S}_t\mathbf{K}_t^T $.\\
		\textbf{Time Update:}\\
		Compute the predicted state $\hat{\mathbf{x}}_{t+1|t}^{\textrm{e}}$ according to $\hat{\mathbf{x}}_{t+1|t}^{\textrm{e}} =\mathcal{A}\hat{\mathbf{x}}_{t|t}^{\textrm{e}}+ \mathcal{B}\mathbf{u}_{t} $.\\
		Compute the covariance matrix $\Sigma_{t+1|t}^{\textrm{e}}$ according to $\Sigma_{t+1|t}^{\textrm{e}} = \mathcal{Q} + \mathcal{A}_t\Sigma_{t|t}^{\textrm{e}}\mathcal{A}_t^T$.
	}
	\SetKw{Fin}{Output:}
	\Fin{} The PDF $p(\mathbf{x}_t^{\textrm{e}}|\mathbf{y}_{1:t})\sim \mathcal{N}\left(\mathbf{x}_t^{\textrm{e}}; \hat{\mathbf{x}}_{t|t}^{\textrm{e}},\Sigma_{t|t}^{\textrm{e}}\right) $ and the PDF $p(\mathbf{x}_{t+1}^{\textrm{e}}|\mathbf{y}_{1:t})\sim \mathcal{N}\left(\mathbf{x}_{t+1}^{\textrm{e}}; \hat{\mathbf{x}}_{t+1|t}^{\textrm{e}},\Sigma_{t+1|t}^{\textrm{e}}\right)$ for $t=1,\dots,N$.
	\caption{Unscented Kalman Filter} 
\end{algorithm}
\subsection{Quantized Kalman Filtering and Smoothing}
The quantized Kalman Filter is an alternative version of the Kalman filter (KF) that modified the measurement update equation to include the quantization effect in the computation of the filtering distributions. This modification is performed in some manners as is shown in \cite{Gomez2020,leong2013}. In this work, we used the following modification of the KF:
\begin{align}\label{eqn:modification_quantized_kalman}
	\hat{\mathbf{x}}_{t|t} = \hat{\mathbf{x}}_{t|t-1} + \mathbf{K}_t\left(y_t-\mathfrak{q}\left\lbrace \mathbf{C}\hat{\mathbf{x}}_{t|t-1}-\mathbf{D}\mathbf{u}_t\right\rbrace\right),
\end{align}
where $\mathbf{K}_t$ is the Kalman gain in the KF. Notice that, with the modification in \eqref{eqn:modification_quantized_kalman}, the QKS algorithm is similar to the standard Kalman smoother (KS). 
\subsection{Gaussian Sum Filtering and Smoothing}
The Gaussian Sum Filter \cite{Cedeno2021,Cedeno2021b} is a novel approach to deal with quantized output data. The key idea of GSF is to approximate the integral of $p(y_t|\mathbf{x}_t)$  given in \eqref{eqn:integral_thm_pytxt} using the Gauss--Legendre quadrature rule. This approximation produces a model with Gaussian sum structure as follows:
\begin{equation}\label{eqn:app_thm_pytxtbygmm}
	p(y_t|\mathbf{x}_t) \approx \sum_{\tau=1}^{K}\varsigma_{t}^{\tau}\mathcal{N}\left(\eta_{t}^{\tau}; \mathbf{C}\mathbf{x}_t+\mathbf{D}\mathbf{u}_t+\bm{\mu}_{t}^{\tau},\mathbf{R}\right), 
\end{equation}
where $K$ is the number of points from the Gauss--Legendre quadrature rule, $\varsigma_{t}^{\tau}$, $\eta_{t}^{\tau}$, and $\bm{\mu}_{t}^{\tau}$ are defined in Table~\ref{tab:my-table2}, and $\omega_{\tau}$ and $\psi_{\tau}$ are weights and points defined by the quadrature rule, given in, e.g.,~\cite{cohen2011numerical}.
\begin{center}
\begin{table}[!ht]
	\normalsize
	\caption{Parameters of the $p(y_t|\mathbf{x}_t)$ approximation.}
	\label{tab:my-table2}
	\setlength{\tabcolsep}{10pt}
	\setlength{\extrarowheight}{3pt}
	\centerline{
			\begin{tabular}{|c|c|c|c|c|}
				\hline 
				\multirow{4}{*}{FLQ}
				&$y_t$       & $\varsigma_{t}^{\tau}$ & $\eta_{t}^{\tau}$ & $\bm{\mu}_{t}^{\tau}$\\ \cline{2-5} 
				&$\psi_{1}$  & $2\omega_{\tau}/(1+\psi_{\tau})^2$  & $-(1-\psi_{\tau})/(1+\psi_{\tau})$ & $-q_1$\\ \cline{2-5} 
				&$\begin{matrix}
					\psi_k,\\k=2,\dots,L-1
				\end{matrix}$&  $\omega_{\tau}(q_{k}-q_{k-1})/2$ & $\psi_{\tau}(q_{k}-q_{k-1})/2$ &  $-(q_{k}+q_{i-1})/2$\\ \cline{2-5} 
				&$\psi_{L}$ & $2\omega_{\tau}/(1+\psi_{\tau})^2$  & $(1-\psi_{\tau})/(1+\psi_{\tau})$  & $-q_{L-1}$\\ \hline 
				ILQ
				&$\begin{matrix}
					\psi_k,\\ k=\dots,1,\dots,L,\dots
				\end{matrix}$ & $\omega_{\tau}(q_{k}-q_{i-1})/2$ & $\psi_{\tau}(q_{k}-q_{i-1})/2$ &  $-(q_{k}+q_{i-1})/2$ \\ \hline 
		\end{tabular}}
	\end{table}
\end{center}
Utilizing the approximation of $p(y_t|\mathbf{x}_{t})$ given in \eqref{eqn:app_thm_pytxtbygmm} the Gaussian sum filter iterates between the following two steps:
\begin{align}\label{eqn:correction}
	p(\mathbf{x}_t|y_{1:t})&=
	\sum_{k=1}^{M_{t|t}} \gamma_{t|t}^{k}\mathcal{N}(\mathbf{x}_{t};\hat{\mathbf{x}}_{t|t}^{k},\Sigma_{t|t}^{k}),\\
	p(\mathbf{x}_{t+1}|y_{1:t})&=\sum_{k=1}^{M_{t+1|t}} \gamma_{t+1|t}^{k}\mathcal{N}(\mathbf{x}_{t+1};\hat{\mathbf{x}}_{t+1|t}^{k},\Sigma_{t+1|t}^{k}),
\end{align}
where all quantities in this recursion can be computed following Algorithm \ref{alg:filtering}. On the other hand, to compute the smoothing distribution, equation \eqref{eqn:forward_backward_smoothing} is separated in two formulas to avoid the division for a non-Gaussian distribution \cite{kitagawa1994two}. The first formula is the backward recursion that,
utilizing the approximation of $p(y_t|\mathbf{x}_{t})$ given in \eqref{eqn:app_thm_pytxtbygmm}, is defined as follows:
\begin{align}\label{eqn:backward_first}
		p(y_{t+1:N}|\mathbf{x}_t)&=\sum_{k=1}^{S_{t|t+1}}\epsilon_{t|t+1}^{k} \lambda_{t|t+1}^{k}\exp\left\lbrace -\frac{1}{2}\left( \mathbf{x}_{t}^T\mathbf{F}_{t|t+1}^{k}\mathbf{x}_{t}-2\mathbf{G}_{t|t+1}^{k T}\mathbf{x}_{t}+H_{t|t+1}^{k}\right) \right\rbrace ,\\
		p(y_{t:N}|\mathbf{x}_t)&=\sum_{k=1}^{S_{t|t}}\epsilon_{t|t}^{k}\lambda_{t|t}^{k}
		\exp\left\lbrace -\dfrac{1}{2}\left( \mathbf{x}_{t}^T\mathbf{F}_{t|t}^{k}\mathbf{x}_{t}-2\mathbf{G}_{t|t}^{kT}\mathbf{x}_{t}+H_{t|t}^{k}\right) \right\rbrace, 
\end{align}
where all quantities in this recursion can be computed following Algorithm \ref{alg:backward}. The second formula computes the smoothing distribution 
as follows
\begin{equation}\label{eqn:smoothing_1}
	p(\mathbf{x}_t|y_{1:N})=\sum_{k=1}^{S_{t|N}}\epsilon_{t|N}^{k}\mathcal{N}(\mathbf{x}_t;\hat{\mathbf{x}}_{t|N}^{k},\Sigma_{t|N}^{k}),
\end{equation}
where all quantities in this equation can be computed following Algorithm \ref{alg:smoothing}.

\begin{algorithm}
	\SetKw{Inicio}{Input:}
	\Inicio{} The PDF of the initial state $p(\mathbf{x}_1)$, e.g.,~$M_{1|0}=1$, $\gamma_{1|0}=1$, $\hat{\mathbf{x}}_{1|0}=\bm{\mu}_1$, $\Sigma_{1|0}=\mathbf{P}_1$. The~points of the Gauss--Legendre quadrature $\left\lbrace \omega_{\tau},\psi_{\tau}\right\rbrace_{\tau=1}^K$.\\
	\For{$t=1$ \textrm{\textbf{to}} $N$}{
		Compute and store $\varsigma_{t}^{\tau}$, $\eta_{t}^{\tau}$, and~$\bm{\mu}_{t}^{\tau}$ according to table \ref{tab:my-table2}.\\
		\textbf{Measurement Update}:\\
		Set $M_{t|t}=KM_{t|t-1}$.\\
		\For{$\ell=1$ \textrm{\textbf{to}} $M_{t|t-1}$}{
			\For{$\tau=1$ \textrm{\textbf{to}} $K$}{
				Compute the index $k=(\ell-1)K+\tau$.\\
				Compute and store the weights, means, and covariances matrices as follows
				\begin{equation}\nonumber
					\begin{split}
						\gamma_{t|t}^{k} &= \bar{\gamma}_{t|t}^{k}\textstyle\left( \sum_{s=1}^{M_{t|t}}\bar{\gamma}_{t|t}^{s}\right)^{-1},\\
						\hat{\mathbf{x}}_{t|t}^{k}&=\hat{\mathbf{x}}_{t|t-1}^{\ell}+\mathbf{K}_t^{\ell}(\eta_{t}^{\tau}-\kappa_t^{\ell\tau}),\\
						\Sigma_{t|t}^{k}&=(\mathbf{I}-\mathbf{K}_t^{\ell}\mathbf{C})\Sigma_{t|t-1}^{\ell},
					\end{split}
					\qquad\qquad
					\begin{split}
						\bar{\gamma}_{t|t}^{k}&=\varsigma_{t}^{\tau} \gamma_{t|t-1}^{\ell}\mathcal{N}_{\eta_{t}^{\tau}}\left( \kappa_t^{ \ell\tau},V_t^{\ell}\right),\\
						\mathbf{K}_t^{\ell}&=\Sigma_{t|t-1}^{\ell}\mathbf{C}^T\left(V_t^{\ell}\right)^{-1}, \\
						\kappa_t^{\ell\tau}&=\mathbf{C}\hat{\mathbf{x}}_{t|t-1}^{\ell}+\mathbf{D}\mathbf{u}_t+\bm{\mu}_{t}^{\tau}, \qquad V_t^{\ell}=R+\mathbf{C}\Sigma_{t|t-1}^{\ell}\mathbf{C}^T.
					\end{split}	
				\end{equation}
			}
		}
		Perform the Gaussian sum reduction algorithm according to~\cite{balenzuela2019} to obtain the reduced GMM of $p(\mathbf{x}_t|y_{1:t})$. \\
		\textbf{Time Update}\\
		Set $M_{t+1|t}=M_{t|t}$.\\
		\For{$k=1$ \textrm{\textbf{to}} $M_{t+1|t}$}{
			Compute and store the weights, means, and covariances matrices as follows:
			\begin{align*}
				\gamma_{t+1|t}^{k}=\gamma_{t|t}^{k}, \qquad \qquad
				\hat{\mathbf{x}}_{t+1|t}^{k}=\mathbf{A}\hat{\mathbf{x}}_{t|t}^{k}+\mathbf{B}\mathbf{u}_t,  \qquad \qquad	\Sigma_{t+1|t}^{k}=\mathbf{Q}+\mathbf{A}\Sigma_{t|t}^{k}\mathbf{A}^T.
			\end{align*}
		}
	}
	\SetKw{Fin}{Output:}
	\Fin{} The~filtering PDFs $p( \mathbf{x}_t|y_{1:t})$, the~predictive PDFs $p(\mathbf{x}_{t+1}|y_{1:t})$, and~the set $\left\lbrace \varsigma_{t}^{\tau}, \eta_{t}^{\tau},\bm{\mu}_{t}^{\tau}\right\rbrace$, for~$t=1,\dots,N$.
	\caption{Gaussian sum filter algorithm for quantized output~data.} 
	\label{alg:filtering}
\end{algorithm}

\begin{algorithm}
	\SetKw{Inicio}{Input:}
	\Inicio{} The initial backward measurement $p(y_N|x_N)$ at $t=N$ with parameters: $S_{N|N}=K$ and
	\begin{equation}\nonumber 
		\begin{split}
			\epsilon_{N|N}^{k}&=\varsigma_N^{k},\\
			\mathbf{F}_{N|N}^{k}&=\mathbf{C}^TR^{-1}\mathbf{C},
		\end{split}
		\qquad \qquad 
		\begin{split}
			\lambda_{N|N}^{k}&=\left(\det\left\lbrace 2\pi R\right\rbrace\right)^{-1/2},\\
			\mathbf{G}_{N|N}^{kT}&=\theta_{N}^{kT}R^{-1}\mathbf{C},
		\end{split}
		\qquad \qquad 
		\begin{split}
			\theta_{N}^{k}&=\eta_{N}^{k}-\mathbf{D}\mathbf{u}_N-\bm{\mu}_N^{k},\\
			H_{N|N}^{k}&=\theta_{N}^{kT}R^{-1}\theta_{N}^{k}.
		\end{split}
	\end{equation}
	The~set $\left\lbrace \varsigma_{t}^{\tau}, \eta_{t}^{\tau},\mu_{t}^{\tau}\right\rbrace$ for $t=1,\dots,N$ computed in algorithm \ref{alg:filtering}.\\
	\For{$t=N-1$ \textrm{\textbf{to}} $1$}{
		\textbf{Backward Prediction}\\
		Set $S_{t|t+1}=S_{t+1|t+1}$.\\ 
		\For{$k=1$ \textrm{\textbf{to}} $S_{t|t+1}$}{
			Compute and store the backward prediction update quantities as follows
			\begin{align*}
				\epsilon_{t|t+1}^{k}&=\epsilon_{t+1|t+1}^{k}, \\
				\lambda_{t|t+1}^{k}&=\left(\det\left\lbrace \mathbf{Q}\right\rbrace \det\left\lbrace \mathbf{F}_{qk}\right\rbrace\right)^{-1/2} \lambda_{t+1|t+1}^{k}, \qquad \qquad \quad \mathbf{F}_{qk}=\mathbf{F}_{t+1|t+1}^{k}+\mathbf{Q}^{-1}\\
				\mathbf{F}_{t|t+1}^{k}&=\mathbf{A}^T\mathbf{M}_{qk}\mathbf{A},  \hspace{51.5mm} \mathbf{M}_{qk}=\mathbf{Q}^{-1}+\mathbf{Q}^{-1}\mathbf{F}_{qk}^{-1}\mathbf{Q}^{-1}\\
				\mathbf{G}_{t|t+1}^{kT}&=\mathbf{G}_{t+1|t+1}^{kT}\mathbf{F}_{qk}^{-1}\mathbf{Q}^{-1}\mathbf{A}-\mathbf{u}_t^T\mathbf{B}^T\mathbf{M}_{qk}\mathbf{A},\\
				H_{t|t+1}^{k}&=H_{t+1|t+1}^{k}-\mathbf{G}_{t+1|t+1}^{kT}\mathbf{F}_{qk}^{-1}\mathbf{G}_{t+1|t+1}^{k} +\mathbf{u}_t^T\mathbf{B}^T\mathbf{M}_{qk}\mathbf{B}\mathbf{u}_t-2\mathbf{u}_t^T\mathbf{B}^T\mathbf{Q}^{-1}\mathbf{F}_{qk}^{-1}\mathbf{G}_{t+1|t+1}^{k}.	
			\end{align*}
		}
		\textbf{Backward Measurement Update}:\\
		Set $S_{t|t}=KS_{t|t+1}$.\\ 
		\For{$\ell=1$ \textrm{\textbf{to}} $S_{t|t+1}$}{
			\For{$\tau=1$ \textrm{\textbf{to}} $K$}{
				Compute the index $k=(\ell-1)K+\tau$.\\
				Compute and store the backward measurement update quantities as follows
				\begin{equation}\nonumber 
					\begin{split}
						\epsilon_{t|t}^{k}&=\varsigma_{t}^{\tau}\epsilon_{t|t+1}^{\ell}, \\
						\mathbf{F}_{t|t}^{k}&=\mathbf{F}_{t|t+1}^{\ell}+\mathbf{C}^TR^{-1}\mathbf{C},						
					\end{split}
					\qquad \qquad
					\begin{split}
						\theta_{t}^{\tau}&=\eta_{t}^{\tau}-Du_{t}-\mu_t^{\tau},\\
						\mathbf{G}_{t|t}^{kT}&=\mathbf{G}_{t|t+1}^{\ell T}+\theta_{t}^{\tau T}R^{-1}\mathbf{C},
					\end{split}
					\qquad \qquad
					\begin{split}
						\lambda_{t|t}^{k}&=\left(\det\left\lbrace 2\pi R\right\rbrace\right)^{-1/2} \lambda_{t|t+1}^{\ell},	\\
						H_{t|t}^{k}&=H_{t|t+1}^{\ell}+\theta_{t}^{\tau T}R^{-1}\theta_{t}^{\tau}.
					\end{split}
				\end{equation}
			}
		}
		Compute the GMM structure of $p(y_{t:N}|\mathbf{x}_t)$ using Lemma A.3 in \cite{Cedeno2021b}.\\ 
		Perform the Gaussian sum reduction algorithm according to~\cite{balenzuela2019} to obtain the reduced GMM structure of $p(y_{t:N}|\mathbf{x}_t)$, see equation (54) in \cite{Cedeno2021b} .\\
		Compute and store the backward filter form of the reduced version of $p(y_{t:N}|\mathbf{x}_t)$ using Lemma A.3 in \cite{Cedeno2021b}. 
	}
	\SetKw{Fin}{Output:}
	\Fin{} The backward prediction $p(y_{t+1:N}|\mathbf{x}_t)$ and~the backward measurement update $p(y_{t:N}|\mathbf{x}_t)$ for $t=N,\dots,1$.
	\caption{Backward-filtering algorithm for quantized output~data.} 
	\label{alg:backward}
\end{algorithm}

\begin{algorithm}
	\SetKw{Inicio}{Input:}
	\Inicio{} The PDFs $p(\mathbf{x}_t|y_{1:t-1})$ and $p(\mathbf{x}_N|y_{1:N})$ obtained from Algorithm \ref{alg:filtering} and the reduce version of $p(y_{t:N}|\mathbf{x}_t)$ obtained from Algorithm \ref{alg:backward}, see equation (54) in \cite{Cedeno2021b}.\\
	Save the PDF $p(\mathbf{x}_N|y_{1:N})$.\\
	\For{$t=N-1$ \textrm{\textbf{to}} $1$}{
		Set $S_{t|N}=M_{t|t-1}S_{\textrm{red}}$.\\
		\For{$\ell=1$ \textrm{\textbf{to}} $S_{\textrm{red}}$}{
			\For{$\tau=1$ \textrm{\textbf{to}} $M_{t|t-1}$}{
				Compute the index $k=(\ell-1)M_{t|t-1}+\tau$.\\
				Compute and store the weights, means, and covariances matrices as follows
				\begin{equation}\nonumber
					\begin{split}
						\epsilon_{t|N}^{k}&=\bar{\epsilon}_{t|N}^{\,k}\left(\textstyle\sum_{s=1}^{S_{t|N}}\bar{\epsilon}_{t|N}^{\,s}\right)^{-1},\\
						\hat{\mathbf{x}}_{t|N}^{k}&=(\mathbf{L}_t^{\ell\tau})^{-1}\rho_t^{\ell\tau},\\
						\Sigma_{t|N}^{k}&=(\mathbf{L}_t^{\ell\tau})^{-1},\\
					\end{split}
					\qquad\qquad
					\begin{split}
						\mathbf{L}_t^{\ell\tau}&=\left[ \left(\mathbf{U}_{t|t}^{\ell}\right)^{-1}+\left(\Sigma_{t|t-1}^{\tau}\right)^{-1}\right],\\
						\rho_t^{\ell\tau}&=\left[\left( \mathbf{U}_{t|t}^{\ell}\right) ^{-1}z_{t|t}^{\ell}+\left( \Sigma_{t|t-1}^{\tau}\right) ^{-1}\hat{\mathbf{x}}_{t|t-1}^{\tau}\right],
					\end{split}
				\end{equation} 
				and
				\begin{equation}\nonumber
					\bar{\epsilon}_{t|N}^{\,k}=\gamma_{t|t-1}^{\tau}\delta_{t|t}^{\ell}\exp\left\lbrace -\dfrac{1}{2}\left[ \phi_{1t}^{\ell}+\phi_{2t}^{\tau}-\phi_{3t}^{\ell\tau}\right] \right\rbrace\left( (2\pi)^{\frac{n}{2}}\sqrt{\det\left\lbrace \mathbf{L}_t^{\ell\tau}\right\rbrace \det\left\lbrace \mathbf{U}_{t|t}^{\ell}\right\rbrace \det\left\lbrace\Sigma_{t|t-1}^{\tau}\right\rbrace}\right)^{-1},
				\end{equation}
			\begin{equation}\nonumber
					\phi_{1t}^{\ell}=\left( z_{t|t}^{\ell}\right)^T\left(\mathbf{U}_{t|t}^{\ell}\right)^{-1}\left(z_{t|t}^{\ell}\right),\quad
					\phi_{2t}^{\tau}=\left(\hat{\mathbf{x}}_{t|t-1}^{\tau}\right)^T\left(\Sigma_{t|t-1}^{\tau}\right)^{-1}\left(\hat{\mathbf{x}}_{t|t-1}^{\tau}\right),\quad
					\phi_{3t}^{\ell\tau}=\left(\rho_{t}^{\ell\tau}\right)^T\left(\mathbf{L}_t^{\ell\tau}\right)^{-1}\left(\rho_{t}^{\ell\tau}\right).
			\end{equation}
			} 
		}
	}
	\SetKw{Fin}{Output:}
	\Fin{}The smoothing PDFs $p(x_t|y_{1:N})$, for~$t=1,\dots,N$.
	\caption{Gaussian sum smoothing algorithm for quantized output~data.} 
	\label{alg:smoothing}
\end{algorithm}

\subsection{Particle Filtering and Smoothing}
Particle filtering \cite{Doucet2000,Gordon1993} is a Monte Carlo method that approximately represents the filtering distributions $p(\mathbf{x}_{t}|y_{1:t})$ of the state variables conditioned to the observations $y_{1:t}$ by using a set of weighted random samples called particles so that:
\begin{equation}
	p(\mathbf{x}_{t}|y_{1:t}) \approx \sum_{i=1}^{M} w_t^{(i)} \delta\left( \mathbf{x}_t-\mathbf{x}_t^{(i)}\right), 
\end{equation}
where $\delta\left(\cdot\right)$ is the Dirac delta function, $w_t^{(i)}$ denotes the $i$th weight, $\mathbf{x}_t^{(i)}$ denotes the $i$th particle sampled from the filtering distribution $p(\mathbf{x}_{t}|y_{1:t})$, and $M$ is the number of particles. Since the filtering distribution is unknown in the current iteration, it is difficult or impossible to sample directly from it. In this case, the particles are usually generated from a known density that is chosen (by the user) to facilitate the sampling process. This is called importance sampling and the PDF is called importance density. Then, the importance weight computation can be carried out in recursive fashion (Sequential importance sampling, SIS) as follows:
\begin{equation}\label{eqn:filter_weights}
	w_t^{(i)}  \propto w_{t-1}^{(i)} \dfrac{p(y_t|\mathbf{x}_t^{(i)})p(\mathbf{x}_t^{(i)}|\mathbf{x}_{t-1}^{(i)})}{h(\mathbf{x}_t|\mathbf{x}^{(i)}_{t-1},y_t)},
\end{equation}
where $h(\mathbf{x}_t|\mathbf{x}_{t-1}^{(i)},y_t)$ is the importance density and $w_{t-1}^{(i)}$ are the importance weights of the previous iteration. On the other hand, the choice of importance distribution is critical for performing the particle filtering and smoothing. The particle filter literature shows that the importance density $p(\mathbf{x}_t|\mathbf{x}^{(i)}_{t-1},y_t)$ is  optimal in the sense that it minimizes the variance of the importance weights $w_t^{(i)}$ \cite{Doucet2000,sarkka2013bayesian}. However, in most cases it is difficult or impossible to draw samples for this optimal importance density, except for particular cases such as a state-space model with nonlinear process and linear output equation \cite{Doucet2000}. Many sub-optimal methods have been developed to approximate the importance density such as Markov Chain Monte Carlo \cite{Liu1998}, ensemble Kalman filter \cite{Bi2015}, local linearization of the state space model, local linearization of the optimal importance distribution \cite{Doucet2000}, among others. One of the most commonly used importance densities in the literature is the state transition prior $p(\mathbf{x}_{t}|\mathbf{x}_{t-1})$, see e.g. \cite{hostettler2015,Cedeno2021,Doucet2000}. This choice yields a intuitive and simple to implement algorithm with $w_t^{(i)} \propto w_{t-1}^{(i)} p(y_t|\mathbf{x}_t^{(i)})$.  This algorithm is called \textit{bootstrap} filter \cite{Gordon1993}.

The particle filter suffers from an impossible to avoid problem called \textit{degeneracy phenomenon}. As shown in \cite{Doucet2000},  the variance of the importance weights can only increase over time. This implies that after a few iterations, most particles have negligible weights, see also \cite{Arulampalam2002}. A consequence of the degeneracy problem is that a large computational effort is devoted to updating particles whose contribution to the final estimate is nearly zero. To solve the degeneracy problem, the \textit{resampling approach} was proposed in \cite{Gordon1993}. The resampling method eliminates the particles that have small weights and the particles with large weights are replicated, generating a new set (with replacement) of equally-weighted particles.    

Additionally, the resampling method used to reduce the degeneracy effect on the particles produces other unwanted issue called \textit{particle impoverishment}. This means a loss of diversity in the sample set since the resampled particles will contain many repeated points that were generated from heavily weighted particles. In the worst-case scenario, all particles can be produced by a single particle with a large weight \cite{Zhai2004}. To solve the impoverishment problem, some methods such as roughening and regularization have been suggested in the literature \cite{Li2015b}. Markov Chain Monte Carlo (MCMC) technique is another method used after the resampling step to add variability to the resampled particles \cite{doucet2001}. The basic idea is to apply the MCMC algorithm to each resampled particle with $p(\mathbf{x}_t|y_{1:t})$ as the target distribution. That is, we need to build a Markov chain by sampling a proposal particle $\mathbf{x}_t^{*}$ from the proposal density. Then, $\mathbf{x}_t^{*}$ is accepted only if $u\leq \varpi(\mathbf{x}_t^{*},\mathbf{x}_t)$, with $u \sim \mathcal{U}[0,1]$, where $\mathcal{U}[a_1,a_2]$ correspond to the uniform distribution over the real numbers in the interval $[a_1,a_2]$, and $\varpi(\mathbf{x}_t^{*},\mathbf{x}_t)$ is the acceptance ratio given by
\begin{equation}\label{eqn:acceptance_ratio}
	\varpi(\mathbf{x}_t^{*},\mathbf{x}_t)=\min\left\lbrace 1,\dfrac{p(y_t|\mathbf{x}_t^{*})}{p(y_t|\mathbf{x}_t)}\right\rbrace,
\end{equation} 
With this process, the diversity of the new particles is greater than the resampled particles, reducing the risk of particle impoverishment. Additionally, the new particles are distributed according to $p(\mathbf{x}_t|y_{1:t})$, as desired. In this paper, we used the MH and RWM algorithms to build the MCMC step. In Algorithm \ref{alg:particle_filter} we summarized the steps to implement the particle filter with the MCMC step. 
\begin{algorithm}
	\label{alg:particle_filter}
	\SetKw{Inicio}{Input:}
	\Inicio{}  $p(\mathbf{x}_1)$, the number of particles $M$. \\
	Draw the samples $\mathbf{x}_1^{(i)} \sim p(\mathbf{x}_1)$ and set $w_1^{(i)}=1/M$ for $i=1,\dots,M$.\\
	\For{$t=1$ \textrm{\textbf{to}} $N$}{
		From the importance distribution draw the samples  $\mathbf{x}_t^{(i)} \sim h(\mathbf{x}_t|\mathbf{x}_{t-1}^{(i)},y_t)$ for $i=1,\dots,M$.  \\
		Calculate the weights $w_t^{(i)}$ using $p(y_t|\mathbf{x}_t)$ given in \eqref{eqn:integral_thm_pytxt} according to \eqref{eqn:filter_weights} for $i=1,\dots,M$.\\
		Normalize the weights $w_t^{(i)}$ to sum to unity.\\
		Do resampling\footnote{In the resampling algorithms MR, SR, and MTR $w_t^{(i)}=1/M$ for $i=1,\dots,M$. The LS algorithm produces a new set of weights, see e.g. \cite{Li2015b}.} and generate a new set of weights $w_t^{(i)}$ and particles $\mathbf{x}_t^{(i)}$ for $i=1,\dots,M$.\\
		Do the MCMC move: for $\ell=1,\dots,M$.\\
		\Indp Pick the sample $\mathbf{x}_t^{(\ell)}$ from the set of the resampled particles.\\
		\Indp \textbf{MH:} Sample a proposal particle $\mathbf{x}_t^{*}$ from the proposal PDF.\\
		\textbf{RWM:} Generate $\mathbf{x}_t^{+}$ from $\mathcal{N}(0,\Lambda^2)$ and compute $\mathbf{x}_t^{*}=\mathbf{x}_t^{(\ell)}+\mathbf{x}_t^{+}$.\\
		\Indm Evaluate $\varpi(\mathbf{x}_t^{*},\mathbf{x}_t^{(\ell)})$ given in \eqref{eqn:acceptance_ratio}. If $u\leq\varpi(\mathbf{x}_t^{*},\mathbf{x}_t^{(\ell)})$, then accept the move $\mathbf{x}_t^{(\ell)}=\mathbf{x}_t^{*}$ else reject the move $\mathbf{x}_t^{(\ell)}=\mathbf{x}_t^{(\ell)}$.
	}
	\SetKw{Fin}{Output:}
	\Fin{} $w_t^{(i)}$, and $\mathbf{x}_t^{(i)} \sim p(\mathbf{x}_t|y_{1:t})$, $i=1,\dots,M$.
	\caption{MCMC-based Particle Filter} 
\end{algorithm}

Similar to particle filtering, particle smoothing is a Monte Carlo method that approximately represents the smoothing distributions $p(\mathbf{x}_{t}|y_{1:N})$ of the state variables conditioned to the observations $y_{1:N}$, using random samples as follows:
\begin{equation}\label{eqn:particle_smoother}
	p(\mathbf{x}_{t}|y_{1:N}) \approx \sum_{i=1}^{M} w_{t|N}^{(i)} \delta\left( \mathbf{x}_t-\tilde{\mathbf{x}}_t^{(i)}\right) ,
\end{equation}
where $w_{t|N}^{(i)}$ denotes the $i$th weight, $\tilde{\mathbf{x}}_t^{(i)}$ denotes the $i$th particle sampled from the smoothing distribution $p(\mathbf{x}_{t}|y_{1:N})$, and $M$ is the number of particles. Some smoothing algorithms are based on the particles provided by the particle filtering, i.e. $\mathbf{x}_t^{(i)}$, such as Backward-simulation particle smoother \cite{Godsill2004}, and marginal particle smoother \cite{Doucet2000}. Particularly in the marginal particle smoother, the weights $w_{t|N}^{(i)}$ are updated in time reverse as follows:
\begin{equation}
	w_{t|N}^{(i)} = \sum_{j=1}^{M} w_{t+1|N}^{(j)}\dfrac{w_{t}^{(i)}p(\mathbf{x}_{t+1}^{(j)}|\mathbf{x}_{t}^{(i)})}{\sum_{k=1}^{M}w_{t}^{(k)}p(\mathbf{x}_{t+1}^{(j)}|\mathbf{x}_{t}^{(k)})},
\end{equation}
where $w_{N|N}^{(i)}=w_{N}^{(i)}$ for $i=1,\dots,M$ and the approximation of \eqref{eqn:particle_smoother} is performed using $\tilde{\mathbf{x}}_t^{(i)}=\mathbf{x}_t^{(i)}$ for $t=N,\dots,1$. In this paper we use the smoothing method developed in \cite{douc2011}, which for the problem of interest in this work, admits further simplifications, see also \cite{wills2013}. This smoothing method \cite{douc2011} require the evaluation of the function $f(\mathbf{x}_{t+1}^{(i)},\mathbf{x}_{t}^{(\tau)})$ given by
\begin{equation}
	f(\mathbf{x}_{t+1}^{(i)},\mathbf{x}_{t}^{(\tau)}) = \exp\left\lbrace -\dfrac{1}{2}\bm{\eta}_t^T\mathbf{Q}^{-1}\bm{\eta}_t \right\rbrace ,
\end{equation}
where $\bm{\eta}_t=\mathbf{x}_{t+1}^{(i)}-\mathbf{A}\mathbf{x}_{t}^{(\tau)}-\mathbf{B}\mathbf{u}_t$. In Algorithm \ref{alg:particle_smoother} we summarize the steps to implement the particle smoother.
\begin{algorithm}
	\label{alg:particle_smoother}
	\SetKwRepeat{Do}{do}{while}%
	\SetKw{Inicio}{Input:}
	\Inicio{}  Weights $w_t^{(i)}$, and particles $\mathbf{x}_t^{(i)}$ provided by Algorithm \ref{alg:particle_filter}, for $t=1,\dots,N$, and $i=1,\dots,M$ . \\
	Set $\tilde{\mathbf{x}}_N^{(i)}=\mathbf{x}_N^{(i)}$ and $w_{N|N}^{(i)}=1/M$ for $i=1,\dots,M$.\\
	\For{$t=N-1$ \textrm{\textbf{to}} $1$}{
		\For{$i=1$ \textrm{\textbf{to}} $M$}{
			\Do{$u>f(\mathbf{x}_{t+1}^{(i)},\mathbf{x}_{t}^{(\tau)})$}{
				Take $\tau \sim \mathcal{U}([1, . . . , M])$.\\
				Take $u \sim \mathcal{U}[0,1]$.\\
			}
			Set $\tilde{\mathbf{x}}_{t}^{(i)}=\mathbf{x}_{t}^{(\tau)}$ and 
			$w_{t|N}^{(i)}=1/M$.
		}
	}
	\SetKw{Fin}{Output:}
	\Fin{} $w_{t|N}^{(i)}$, and $\tilde{\mathbf{x}}_{t}^{(i)} \sim p(\mathbf{x}_t|y_{1:N})$, $i=1,\dots,M$.
	\caption{Rejection based Particle Smoother} 
\end{algorithm}
On the other hand, provided the weights $w^{(i)}$ and particles $\mathbf{x}^{(i)}$ (from particle filter or smoother) the state estimators in \eqref{eqn:filtered_state} and \eqref{eqn:smoothed_state} and the covariance matrices of the estimation error in \eqref{eqn:filtered_cov_state} and \eqref{eqn:smoothed_cov_state} can be computed as follows:
\begin{equation}
	\mathbb{E}\left\lbrace g(\mathbf{x}_t)|s\right\rbrace \approx \sum_{i=1}^{M}w^{(i)}g(\mathbf{x}^{(i)}),
\end{equation}
where $g(\mathbf{x}_t)$ represents a function of $\mathbf{x}_t$. To compute the mean and covariance matrix of the filtering ans smoothing distributions $g(\mathbf{x}_t)=\mathbf{x}_t$ and $g(\mathbf{x}_t)=\left(\mathbf{x}_t-\mathbb{E}\left\lbrace \mathbf{x}_t\right\rbrace\right)\left(\mathbf{x}_t-\mathbb{E}\left\lbrace \mathbf{x}_t\right\rbrace\right)^T$, respectively. The variable $s$ represents the observations set used, that for the filtering estimations $s=y_{1:t}$ and for smoothing estimation $s=y_{1:N}$.
\section{Numerical experiment} \label{sec:numerical_experiments}
In this section, we present a numerical example to analyze the performance of KF/KS, EKF/EKS, QKF/QKS, UKF/UKS, GSF/GSS, and MCMC-based PF/PS having quantized observations. We use the discrete-time system in the state-space form given in \eqref{eqn:general_system_state}-\eqref{eqn:general_system_output} with 
\begin{equation}
	y_t=\Delta\textrm{round}\left( z_t/\Delta\right),
\end{equation}
where $\Delta$ is a quantization step and \textit{round} is the Matlab function that computes the nearest decimal or integer. The sets $\mathcal{R}_k$ are computed using $q_{k-1}=y_t-0.5\Delta$ and $q_{k}=y_t+0.5\Delta$. We compare the performance of all filtering and smoothing algorithms considering eight variations of the PF where we use the Markov Chain Monte Carlo method MH and RWM with the following resampling methods: SYS, ML, MT, and LS. For clarity of presentation we use the bootstrap filter and we solve the integral in \eqref{eqn:integral_thm_pytxt} using the cumulative distribution function computed with the Matlab function \textit{mvncdf}. We consider the state-space system given in \eqref{eqn:general_system_state}-\eqref{eqn:general_system_output} with $\mathbf{A}=0.9$, $\mathbf{B}=1.2$, $\mathbf{C}=2.2$, and $\mathbf{D}=0.75$. We also consider that $\mathbf{w}_{t}\sim\mathcal{N}\left(\mathbf{w}_{t};0,1\right)$, $v_{t}\sim\mathcal{N}\left(v_{t};0,0.5\right)$, the input signal is drawn from $\mathcal{N}\left(0,1\right)$, and $\mathbf{x}_1\sim\mathcal{N}(\mathbf{x}_1;1,0.01)$. We consider $\Delta=8$, $K=10$, $\Lambda^2=100$, and $M=\left\lbrace 100,500,1000\right\rbrace $ particles to perform the particle filtering and smoothing. 

\begin{figure*}[!ht]
	\centerline{
		\includegraphics[width=1\linewidth]{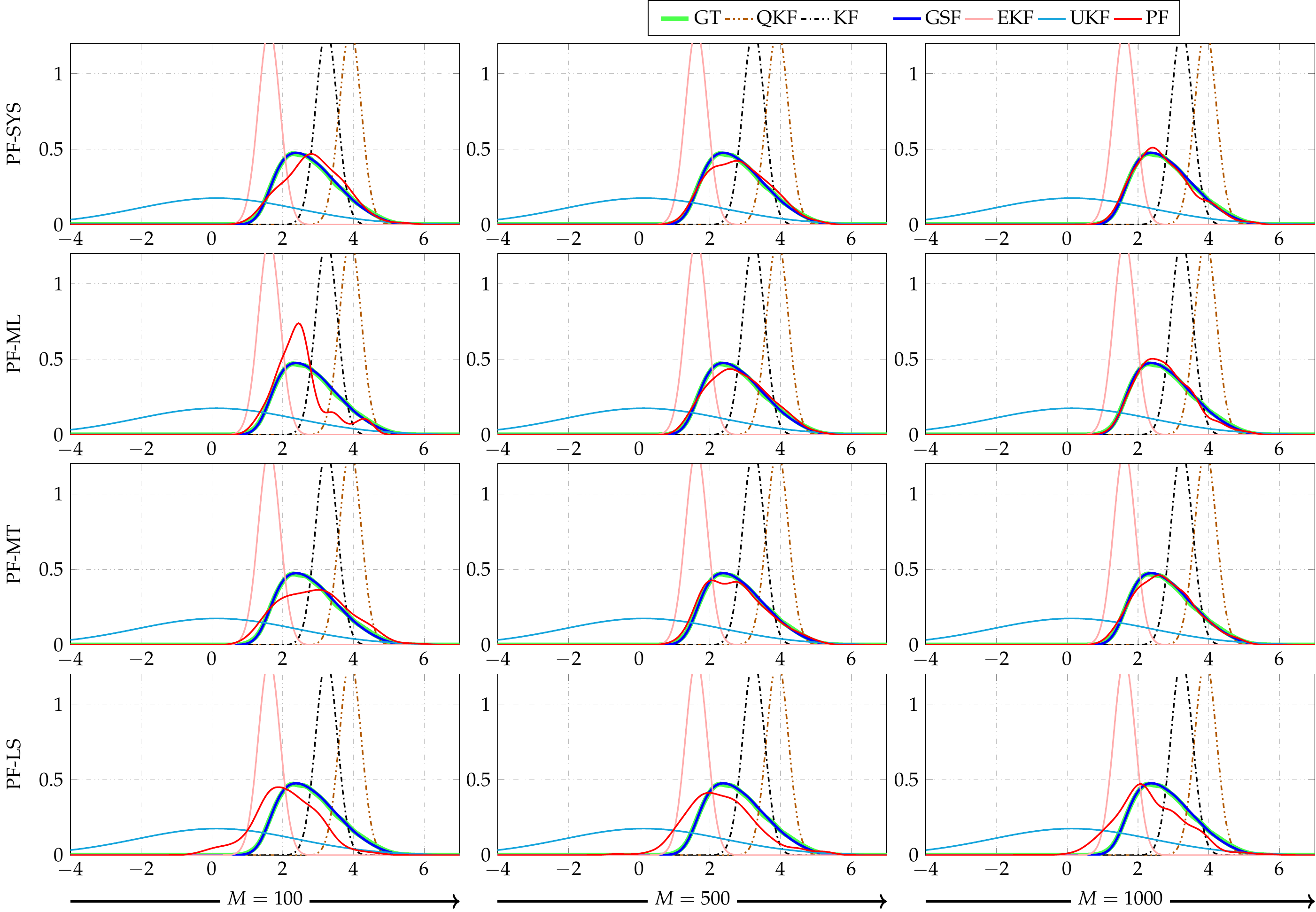}}
	\caption{Filtering PDFs for a time instant. GT stands for the ground truth. KF, EKF, QKF, UKF, GSF, and PF stand for Kalman filter, extended Kalman filter, quantized Kalman filter, unscented Kalman filter, Gaussian sum filter, and particle filter, respectively. The PDFs given by the KF, EKF, QKF, UKF were frozen in all plots to observe the behavior of the PF (with RWM moves) when the number of particles increases. SYS, ML, MT, LS stand for systematic, multinomial, metropolis, and local selection resampling algorithms, respectively. }  
	\label{fig:escalar_filtered_some_instants}                                                               
\end{figure*}

\begin{figure*}[!ht]
	\centerline{
		\includegraphics[width=1\linewidth]{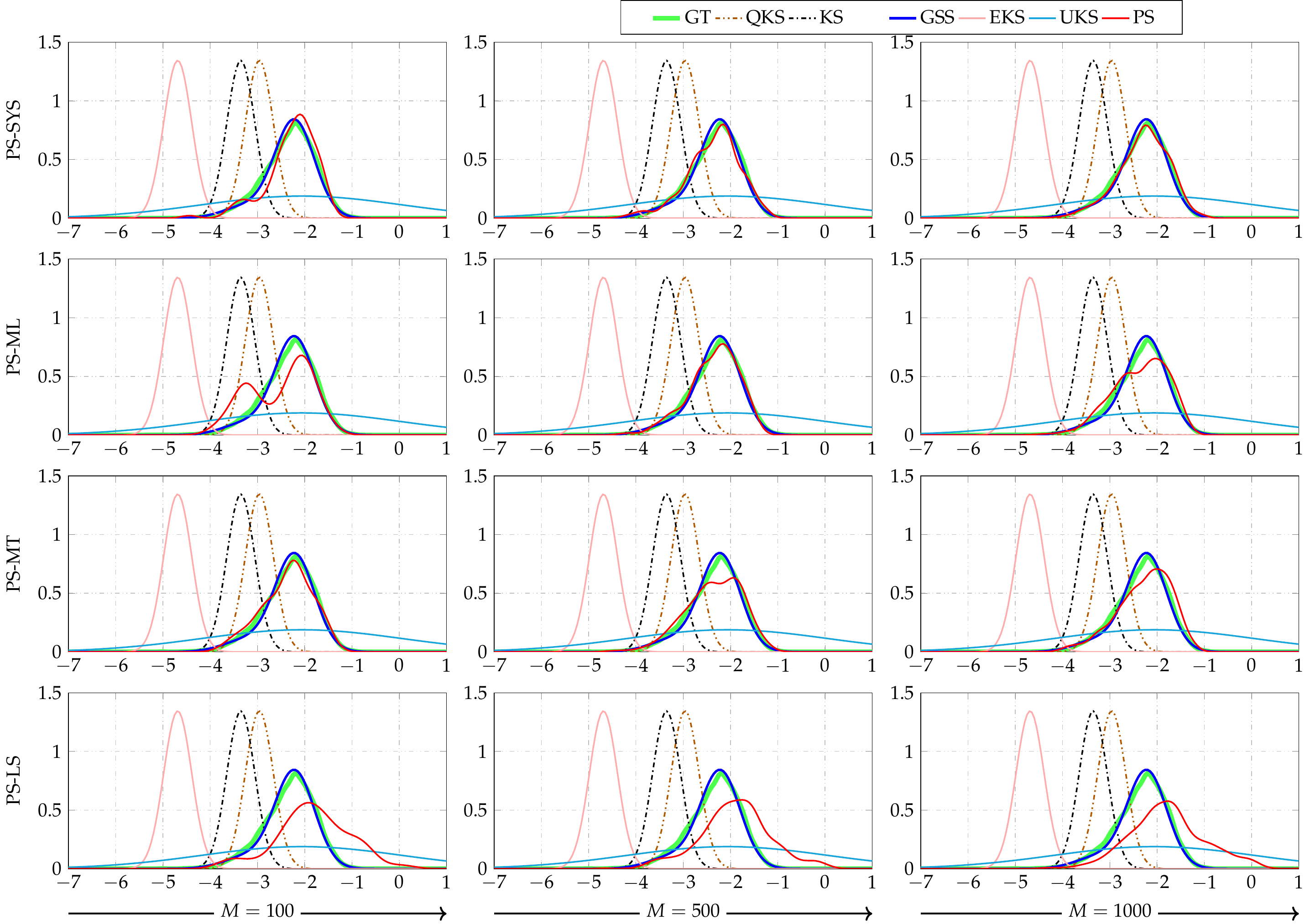}}
	\caption{Smoothing PDFs for a time instant. GT stands for the ground truth. KS, EKS, QKS, UKS, GSS, and PS stand for Kalman smoother, extended Kalman smoother, quantized Kalman smoother, unscented Kalman smoother, Gaussian sum smoother and particle smoother, respectively. The PDFs given by the KS, EKS, QKS, UKS were frozen in all plots to observe the behavior of the PS (with RWM moves) when the number of particles increases. SYS, ML, MT, LS stand for systematic, multinomial, metropolis, and local selection resampling algorithms, respectively.}  
	\label{fig:escalar_smoothed_some_instants}                                                               
\end{figure*}

\begin{figure}[!ht]
	\centerline{
		\includegraphics[width=1\linewidth]{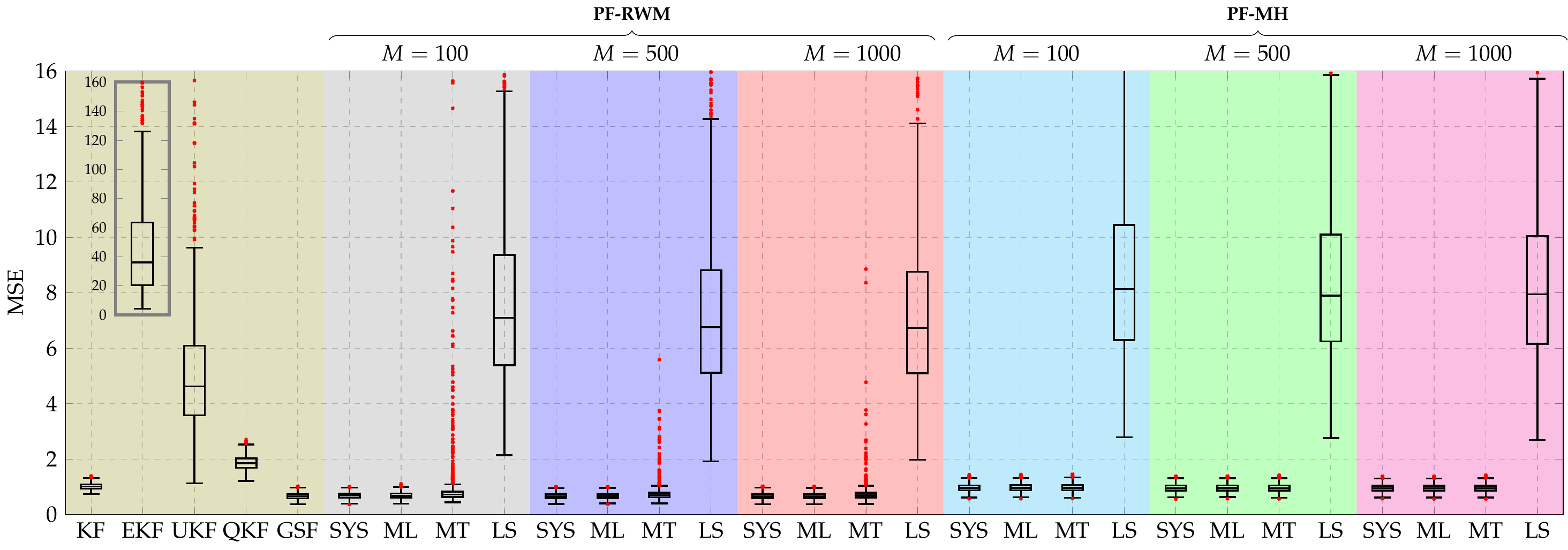}}
	\caption{Boxplot of the MSE between the estimated and true state for 1000 Monte Carlo experiments. KF, EKF, QKF, UKF, GSF, and PF stand for Kalman filter, extended Kalman filter, quantized Kalman filter, unscented Kalman filter, Gaussian sum filter, and particle filter, respectively. Additionally, SYS, ML, MT, LS stand for systematic, multinomial, metropolis, and local selection resampling algorithms, respectively. RWM and MH denote Random-Walk Metropolis and Metropolis-Hasting moves.}  
	\label{fig:escalar_filtered_boxplot_mse}                                                               
\end{figure}

\begin{figure}
	\centerline{
		\includegraphics[width=1\linewidth]{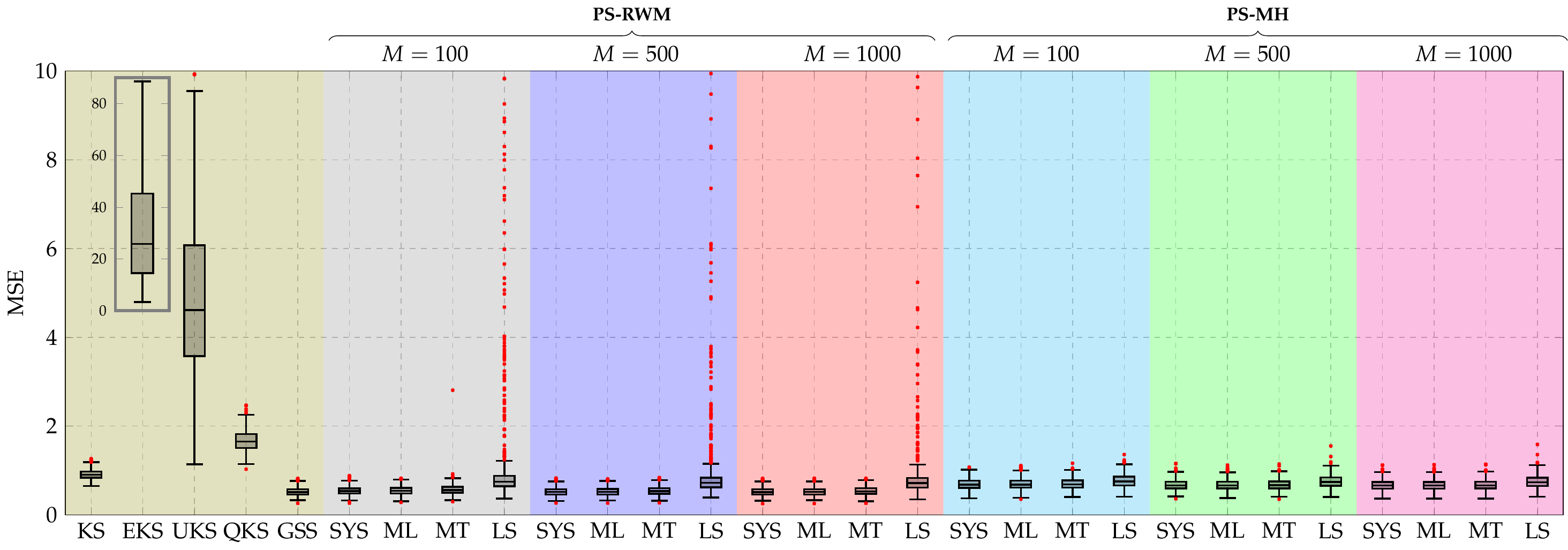}}
	\caption{Boxplot of the MSE between the estimated and true state for 1000 Monte Carlo experiments. KS, EKS, QKS, UKS, GSS, and PS stand for Kalman smoother, extended Kalman smoother, quantized Kalman smoother, unscented Kalman smoother, Gaussian sum smoother, and particle smoother, respectively. Additionally, SYS, ML, MT, LS stand for systematic, multinomial, metropolis, and local selection resampling algorithms, respectively. RWM and MH denote Random-Walk Metropolis and Metropolis-Hasting moves.}  
	\label{fig:escalar_smoothed_boxplot_mse}                                                               
\end{figure}

In Figure \ref{fig:escalar_filtered_some_instants} and \ref{fig:escalar_smoothed_some_instants} we show the filtering and smoothing distributions, i.e. $p(\mathbf{x}_t|y_{1:t})$ and $p(\mathbf{x}_t|y_{1:N})$, for a time instant.  We freeze the results of KF, QKF, EKF, UKF, and GSF to observe the behavior of the PF when varying the number of used samples, and when different MCMC methods are used with different resampling algorithms. These figures show that the PDFs obtained used GSF/GSS are the ones that best fit the ground truth, followed by PF/PS. Furthermore, in Figure \ref{fig:escalar_filtered_boxplot_mse} and \ref{fig:escalar_smoothed_boxplot_mse} we show the boxplot of the mean square error (MSE) between the estimated and the true state running 1000 Monte Carlo experiments. These figures show the loss of accuracy in the state estimation obtained with KF/KS, EKF/EKS, QKF/QKS, and UKF/UKS, and a better performance for GSF/GSS and PF/PS (except from PF version that uses LS resampling). Additionally we can observe that in terms of accuracy, the PF/PS implementation that gives the lower MSE is the one that uses the MCMC move RWM with SYS, ML, and MT resampling methods. On the other hand, in terms of the computational load, PS in all its versions exhibit the most high execution time, followed by the GSS, EKS, UKS and KS. In Table \ref{tab:my-table3} we ranked all the algorithms studied in this work in terms of the mean of the MSE and the execution time. This table suggests that there is a trade off between the accuracy of the estimation and the execution time in the case of PF/PS. The GSF/GSS, on the other hand, exhibits high accuracy in the estimation and a relatively small execution time.

\begin{center}
	\begin{table}[!ht]
		\small
		\caption{Rank of the filtering and smoothing recursive algorithms for quantized data. References: KF/KS, EKF/EKS \cite{sarkka2013bayesian}, QKF/QKS \cite{leong2013,Gomez2020} PF/PS \cite{Gordon1993}, UKF/UKS \cite{Julier1997,sarkka2013bayesian}, GSF/GSS \cite{Cedeno2021,Cedeno2021b}. The notation XX-YY-ZZ(M) denotes the following: XX stands for the filtering or smoothing algorithm, YY stands for the MCMC algorithm, ZZ stands for the resampling method, and (M) stands for the used number of particles.}
		\label{tab:my-table3}
		\centerline{
			\begin{tabular}{ccccccc}
				\toprule\toprule
				\multicolumn{1}{|c|}{\multirow{2}{*}{\textbf{Rank}}} & \multicolumn{2}{c|}{\textbf{Filtering}}  & \multicolumn{2}{c|}{\textbf{Smoothing}}  & \multicolumn{2}{c|}{\textbf{Smoothing Execution Time}} \\ \cmidrule(l){2-7} 
				\multicolumn{1}{|c|}{} & \multicolumn{1}{c|}{\textbf{MSE}} & \multicolumn{1}{c|}{\textbf{Algorithm}} & \multicolumn{1}{c|}{\textbf{MSE}} & \multicolumn{1}{c|}{\textbf{Algorithm}} & \multicolumn{1}{c|}{\textbf{Execution Time}} & \multicolumn{1}{c|}{\textbf{Algorithm}} \\ \bottomrule \bottomrule
				1 &   0.6724 & GSF &   0.5207 & GSS &   0.0026 & KS \\ \hline 
				2 &   0.6740 & PF-RWM-SYS(1000) &   0.5212 & PS-RWM-SYS(1000) &   0.0031 & QKS \\ \hline
				3 &   0.6744 & PF-RWM-ML(1000) &   0.5220 & PS-RWM-ML(1000) &   0.0111 & UKS \\ \hline
				4 &   0.6754 & PF-RWM-SYS(500) &   0.5231 & PS-RWM-SYS(500) &   0.1453 & PS-RWM-SYS(100) \\ \hline
				5 &   0.6765 & PF-RWM-ML(500) &   0.5247 & PS-RWM-ML(500) &   0.1644 & PS-RWM-LS(100) \\ \hline 
				6 &   0.6880 & PF-RWM-SYS(100) &   0.5393 & PS-RWM-MT(1000) &   0.1718 & PS-RWM-ML(100) \\ \hline 
				7 &   0.6948 & PF-RWM-ML(100) &   0.5415 & PS-RWM-SYS(100) &   0.2077 & PS-MH-LS(100) \\ \hline 
				8 &   0.7588 & PF-RWM-MT(1000) &   0.5420 & PS-RWM-MT(500) &   0.2109 & PS-MH-SYS(100) \\ \hline 
				9 &   0.7830 & PF-RWM-MT(500) &   0.5470 & PS-RWM-ML(100) &   0.2354 & PS-MH-ML(100) \\ \hline 
				10 &   0.9590 & PF-MH-SYS(1000) &   0.5689 & PS-RWM-MT(100) &   0.3931 & GSS \\ \hline 
				11 &   0.9593 & PF-MH-MT(1000) &   0.6708 & PS-MH-SYS(1000) &   0.3984 & PS-RWM-MT(100) \\ \hline 
				12 &   0.9595 & PF-MH-ML(1000) &   0.6711 & PS-MH-ML(1000) &   0.4579 & PS-MH-MT(100) \\ \hline 
				13 &   0.9608 & PF-MH-SYS(500) &   0.6737 & PS-MH-SYS(500) &   0.4676 & EKS \\ \hline 
				14 &   0.9612 & PF-MH-MT(500) &   0.6746 & PS-MH-ML(500) &   0.6048 & PS-RWM-SYS(500) \\ \hline 
				15 &   0.9612 & PF-MH-ML(500) &   0.6752 & PS-MH-MT(1000) &   0.6348 & PS-RWM-LS(500) \\ \hline 
				16 &   0.9686 & PF-MH-SYS(100) &   0.6781 & PS-MH-MT(500) &   0.7469 & PS-RWM-ML(500) \\ \hline 
				17 &   0.9697 & PF-MH-ML(100) &   0.6927 & PS-MH-SYS(100) &   0.9772 & PS-MH-LS(500) \\ \hline 
				18 &   0.9715 & PF-MH-MT(100) &   0.6927 & PS-MH-ML(100) &   1.2054 & PS-RWM-SYS(1000) \\ \hline 
				19 &   1.0138 & KF &   0.6974 & PS-MH-MT(100) &   1.2274 & PS-RWM-LS(1000) \\ \hline 
				20 &   1.6731 & PF-RWM-MT(100) &   0.7469 & PS-MH-LS(1000) &   1.2709 & PS-MH-SYS(500) \\ \hline 
				21 &   1.8616 & QKF &   0.7497 & PS-MH-LS(500) &   1.4192 & PS-MH-ML(500) \\ \hline 
				22 &   5.0549 & UKF &   0.7667 & PS-MH-LS(100) &   1.5197 & PS-RWM-ML(1000) \\ \hline 
				23 &   7.3381 & PF-RWM-LS(1000) &   0.9100 & KS &   1.8362 & PS-RWM-MT(500) \\ \hline 
				24 &   7.3602 & PF-RWM-LS(500) &   0.9393 & PS-RWM-LS(1000) &   2.0277 & PS-MH-LS(1000) \\ \hline 
				25 &   7.6912 & PF-RWM-LS(100) &   1.2900 & PS-RWM-LS(500) &   2.3945 & PS-MH-MT(500) \\ \hline 
				26 &   8.3846 & PF-MH-LS(1000) &   1.6693 & QKS &   3.0254 & PS-MH-SYS(1000) \\ \hline 
				27 &   8.4079 & PF-MH-LS(500) &   5.0545 & UKS &   3.3505 & PS-MH-ML(1000) \\ \hline 
				28 &   8.6717 & PF-MH-LS(100) &   6.4904 & PS-RWM-LS(100) &   3.6364 & PS-RWM-MT(1000) \\ \hline 
				29 &  47.7827 & EKF &  33.8842 & EKS &   5.0651 & PS-MH-MT(1000) \\ \bottomrule \bottomrule
		\end{tabular}}
	\end{table}
\end{center}

\begin{figure}
	\centerline{
		\includegraphics[width=1\linewidth]{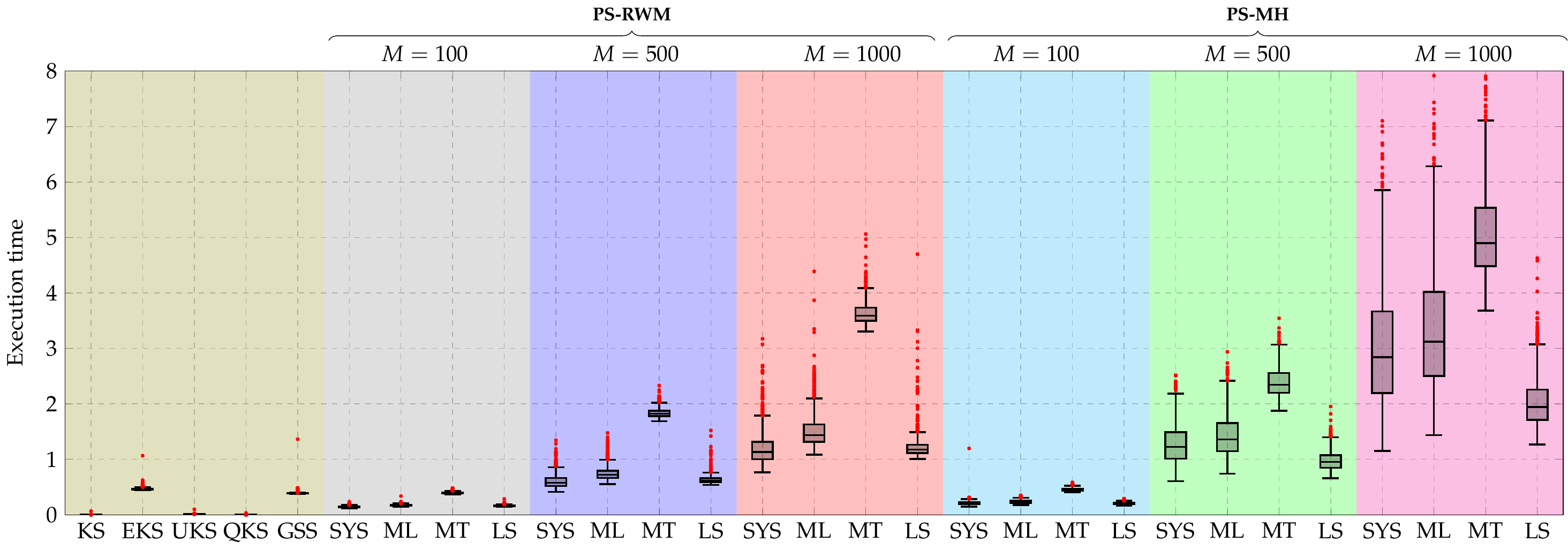}}
	\caption{Boxplot of the execution time for 1000 Monte Carlo experiments. KS, EKS, QKS, UKS, GSS, and PS stand for Kalman smoother, extended Kalman smoother, quantized Kalman smoother, unscented Kalman smoother, Gaussian sum smoother, and particle smoother, respectively. Additionally, SYS, ML, MT, LS stand for systematic, multinomial, metropolis, and local selection resampling algorithms, respectively. RWM and MH denote Random-Walk Metropolis and Metropolis-Hasting moves.}  
	\label{fig:escalar_time_boxplot_mse}                                                               
\end{figure}

\section{Conclusion} \label{sec:conclusions}
In this paper, we investigate the performance of Extended Kalman filter/smoother, Quantized Kalman filter/smoother, Unscented Kalman filter/smoother, Gaussian sum filter/smoother, and particle filter/smoother for state-space models with quantized observations. The analysis was carried out in terms of the accuracy of the estimation, using the Mean square error and the computational cost. The simulations show that the PDFs of Gaussian sum filter/smoother and particle filter/smoother with a high number of particles are the ones that best fit the ground truth PDFs. However, the particle filter/smoother exhibits a high computational load when the number of samples increases. The Extended Kalman filter/smoother, Quantized Kalman filter/smoother, and Unscented Kalman filter/smoother produce results with low accuracy although its execution time is small. From simulations, we observe that the performance of the particle filter is closely related to the choice of the resampling method and the MCMC algorithms, which addresses the degeneracy problem and mitigates the sample impoverishment.  We use four different resampling schemes combined with two MCMC algorithms. We found that the implementation of the MCMC-based particle filter and smoothing that produces the lower MSE is the one that uses Random Walk Metropolis combined with the systematic resampling technique.

\bibliographystyle{ieeetr}
\bibliography{bibliography}

\begin{thebibliography}{10}

\bibitem{gersho2012}
A.~Gersho and R.~M. Gray, {\em {Vector Quantization and Signal Compression}},
  vol.~159.
\newblock Springer Science \& Business Media, 2012.

\bibitem{widrow2008quantization}
B.~Widrow and I.~Koll{\'a}r, {\em Quantization Noise: Roundoff Error in Digital
  Computation, Signal Processing, Control, and Communications}.
\newblock Cambridge University Press, 2008.

\bibitem{Li2011}
S.~Li, D.~Sauter, and B.~Xu, ``{Fault isolation filter for networked control
  system with event-triggered sampling scheme},'' {\em Sensors}, vol.~11,
  no.~1, pp.~557--572, 2011.

\bibitem{Zhang2020}
X.~Zhang, Q.~Han, X.~Ge, D.~Ding, L.~Ding, D.~Yue, and C.~Peng, ``{Networked
  control systems: a survey of trends and techniques},'' {\em IEEE/CAA Journal
  of Automatica Sinica}, vol.~7, no.~1, pp.~1--17, 2020.

\bibitem{Zhang2021}
L.~Zhang, H.~Liang, Y.~Sun, and C.~K. Ahn, ``{Adaptive Event-Triggered Fault
  Detection Scheme for Semi-Markovian Jump Systems With Output Quantization},''
  {\em IEEE Transactions on Systems, Man, and Cybernetics: Systems}, vol.~51,
  no.~4, pp.~2370--2381, 2021.

\bibitem{Noshad2019}
Z.~Noshad, N.~Javaid, T.~Saba, Z.~Wadud, M.~Q. Saleem, M.~E. Alzahrani, and
  O.~E. Sheta, ``{Fault Detection in Wireless Sensor Networks through the
  Random Forest Classifier},'' {\em Sensors}, vol.~19, no.~7, 2019.

\bibitem{Huang2021}
C.~Huang, B.~Shen, L.~Zou, and Y.~Shen, ``{Event-Triggering State and Fault
  Estimation for a Class of Nonlinear Systems Subject to Sensor Saturations},''
  {\em Sensors}, vol.~21, no.~4, 2021.

\bibitem{Liu2020}
S.~Liu, Z.~Wang, J.~Hu, and G.~Wei, ``{Protocol-based extended Kalman filtering
  with quantization effects: The Round-Robin case},'' {\em Int. J. Robust
  Nonlinear Control}, vol.~30, no.~18, pp.~7927--7946, 2020.

\bibitem{Ding2021}
D.~Ding, Q.~L. Han, X.~Ge, and J.~Wang, ``{Secure State Estimation and Control
  of Cyber-Physical Systems: A Survey},'' {\em IEEE Trans. on Systems, Man, and
  Cybernetics: Systems}, vol.~51, no.~1, pp.~176--190, 2021.

\bibitem{Wang2017}
X.~Wang, T.~Li, S.~Sun, and J.~M. Corchado, ``{A Survey of Recent Advances in
  Particle Filters and Remaining Challenges for Multitarget Tracking},'' 2017.

\bibitem{curry1970estimation}
R.~E. Curry, {\em Estimation and control with quantized measurements}.
\newblock MIT press, 1970.

\bibitem{gustafsson2009statistical}
F.~Gustafsson and R.~Karlsson, ``Statistical results for system identification
  based on quantized observations,'' {\em Automatica}, vol.~45, no.~12,
  pp.~2794--2801, 2009.

\bibitem{wang2010system}
L.~Y. Wang, G.~G. Yin, J.~Zhang, and Y.~Zhao, {\em System identification with
  quantized observations}.
\newblock Springer, 2010.

\bibitem{marelli2010scenario}
D.~E. Marelli, B.~I. Godoy, and G.~C. Goodwin, ``A scenario-based approach to
  parameter estimation in state-space models having quantized output data,'' in
  {\em 49th IEEE Conference on Decision and Control (CDC)}, pp.~2011--2016,
  IEEE, 2010.

\bibitem{Rana2015}
M.~M. Rana and L.~Li, ``{An Overview of Distributed Microgrid State Estimation
  and Control for Smart Grids},'' {\em Sensors}, vol.~15, no.~2, 2015.

\bibitem{sarkka2013bayesian}
S.~S{\"a}rkk{\"a}, {\em Bayesian filtering and smoothing}, vol.~3.
\newblock Cambridge University Press, 2013.

\bibitem{anderson2007optimal}
B.~D.~O. Anderson and J.~B. Moore, {\em Optimal control: linear quadratic
  methods}.
\newblock Courier Corporation, 2007.

\bibitem{Julier1997}
S.~J. Julier and J.~K. Uhlmann, ``{New extension of the Kalman filter to
  nonlinear systems},'' in {\em Signal processing, sensor fusion, and target
  recognition VI}, vol.~3068, pp.~182--193, International Society for Optics
  and Photonics, 1997.

\bibitem{arasaratnam2007discrete}
I.~Arasaratnam, S.~Haykin, and R.~J. Elliott, ``{Discrete-time nonlinear
  filtering algorithms using Gauss--Hermite quadrature},'' {\em Proceedings of
  the IEEE}, vol.~95, no.~5, pp.~953--977, 2007.

\bibitem{Sviestins2001}
E.~Sviestins and T.~Wigren, ``{Nonlinear techniques for Mode C climb/descent
  rate estimation in ATC systems},'' {\em IEEE Transactions on Control Systems
  Technology}, vol.~9, no.~1, pp.~163--174, 2001.

\bibitem{Gomez2020}
J.~C. G{\'{o}}mez and G.~D. Sad, ``{A State Observer from Multilevel Quantized
  Outputs},'' in {\em 2020 Argentine Conference on Automatic Control (AADECA)},
  pp.~1--6, 2020.

\bibitem{leong2013}
A.~S. Leong, S.~Dey, and G.~N. Nair, ``{Quantized Filtering Schemes for
  Multi-Sensor Linear State Estimation: Stability and Performance Under High
  Rate Quantization},'' {\em IEEE Transactions on Signal Processing}, vol.~61,
  no.~15, pp.~3852--3865, 2013.

\bibitem{zhou2009}
Y.~Zhou, J.~Li, and D.~Wang, ``{Unscented Kalman Filtering based quantized
  innovation fusion for target tracking in WSN with feedback},'' in {\em 2009
  International Conference on Machine Learning and Cybernetics}, vol.~3,
  pp.~1457--1463, 2009.

\bibitem{Gordon1993}
N.~J. Gordon, D.~J. Salmond, and A.~F.~M. Smith, ``{Novel approach to
  nonlinear/non-Gaussian Bayesian state estimation},'' in {\em IEE proceedings
  F (radar and signal processing)}, vol.~140, pp.~107--113, IET, 1993.

\bibitem{Doucet2000}
A.~Doucet, S.~Godsill, and C.~Andrieu, ``{On sequential Monte Carlo sampling
  methods for Bayesian filtering},'' {\em Statistics and computing}, vol.~10,
  no.~3, pp.~197--208, 2000.

\bibitem{Li2015b}
T.~Li, M.~Bolic, and P.~M. Djuric, ``{Resampling Methods for Particle
  Filtering: Classification, implementation, and strategies},'' {\em IEEE
  Signal Processing Magazine}, vol.~32, no.~3, pp.~70--86, 2015.

\bibitem{Douc2005}
R.~Douc and O.~Cappe, ``{Comparison of resampling schemes for particle
  filtering},'' in {\em ISPA 2005. Proceedings of the 4th International
  Symposium on Image and Signal Processing and Analysis, 2005.}, pp.~64--69,
  2005.

\bibitem{Bi2015}
H.~Bi, J.~Ma, and F.~Wang, ``{An Improved Particle Filter Algorithm Based on
  Ensemble Kalman Filter and Markov Chain Monte Carlo Method},'' {\em IEEE
  Journal of Selected Topics in Applied Earth Observations and Remote Sensing},
  vol.~8, no.~2, pp.~447--459, 2015.

\bibitem{Cedeno2021}
A.~L. Cede{\~{n}}o, R.~Albornoz, R.~Carvajal, B.~I. Godoy, and J.~C.
  Ag{\"{u}}ero, ``{On Filtering Methods for State-Space Systems having Binary
  Output Measurements},'' {\em IFAC-PapersOnLine}, vol.~54, no.~7,
  pp.~815--820, 2021.

\bibitem{Cedeno2021b}
A.~L. Cede{\~{n}}o, R.~Albornoz, R.~Carvajal, B.~I. Godoy, and J.~C.
  Ag{\"{u}}ero, ``{A Two-Filter Approach for State Estimation Utilizing
  Quantized Output Data},'' {\em Sensors}, vol.~21, no.~22, p.~7675, 2021.

\bibitem{Zhai2004}
Y.~Zhai and M.~Yeary, ``{Implementing particle filters with Metropolis-Hastings
  algorithms},'' in {\em Region 5 Conference: Annual Technical and Leadership
  Workshop, 2004}, pp.~149--152, 2004.

\bibitem{Sherlock2015}
C.~Sherlock, A.~H. Thiery, G.~O. Roberts, and J.~S. Rosenthal, ``{On the
  efficiency of pseudo-marginal random walk Metropolis algorithms},'' {\em The
  Annals of Statistics}, vol.~43, pp.~238--275, feb 2015.

\bibitem{degroot2005}
M.~H. DeGroot, {\em {Optimal Statistical Decisions}}.
\newblock Wiley Classics Library, Wiley, 2005.

\bibitem{Gelb1974}
A.~Gelb, J.~Kasper, R.~Nash, C.~Price, and A.~Sutherland, {\em {Applied optimal
  estimation}}.
\newblock Cambridge, MA: MIT Press, 1974.

\bibitem{Grewal2014}
M.~S. Grewal and A.~P. Andrews, {\em {Kalman filtering: Theory and Practice
  with MATLAB}}.
\newblock John Wiley \& Sons, 2014.

\bibitem{Traore2014}
N.~Traor{\'{e}}, L.~{Le Pourhiet}, J.~Frelat, F.~Rolandone, and B.~Meyer,
  ``{Does interseismic strain localization near strike-slip faults result from
  boundary conditions or rheological structure?},'' {\em Geophysical Journal
  International}, vol.~197, pp.~50--62, apr 2014.

\bibitem{Wan2000}
E.~A. Wan and R.~V.~D. Merwe, ``{The unscented Kalman filter for nonlinear
  estimation},'' in {\em Proceedings of the IEEE 2000 Adaptive Systems for
  Signal Processing, Communications, and Control Symposium (Cat. No.00EX373)},
  pp.~153--158, 2000.

\bibitem{cohen2011numerical}
H.~Cohen, {\em Numerical approximation methods}.
\newblock Springer, 2011.

\bibitem{kitagawa1994two}
G.~Kitagawa, ``The two-filter formula for smoothing and an implementation of
  the {G}aussian-sum smoother,'' {\em Annals of the Institute of Statistical
  Mathematics}, vol.~46, no.~4, pp.~605--623, 1994.

\bibitem{balenzuela2019}
M.~P. Balenzuela, J.~Dahlin, N.~Bartlett, A.~G. Wills, C.~Renton, and
  B.~Ninness, ``Accurate gaussian mixture model smoothing using a two-filter
  approach,'' in {\em 2018 IEEE Conference on Decision and Control (CDC)},
  pp.~694--699, 2018.

\bibitem{Liu1998}
J.~S. Liu and R.~Chen, ``{Sequential Monte Carlo Methods for Dynamic
  Systems},'' {\em Journal of the American Statistical Association}, vol.~93,
  pp.~1032--1044, sep 1998.

\bibitem{hostettler2015}
R.~Hostettler, ``{A two filter particle smoother for Wiener state-space
  systems},'' {\em 2015 IEEE Conference on Control and Applications, CCA 2015 -
  Proceedings}, no.~September, pp.~412--417, 2015.

\bibitem{Arulampalam2002}
M.~S. Arulampalam, S.~Maskell, N.~Gordon, and T.~Clapp, ``{A tutorial on
  particle filters for online nonlinear/non-Gaussian Bayesian tracking},'' {\em
  IEEE Trans. on Signal Processing}, vol.~50, no.~2, pp.~174--188, 2002.

\bibitem{doucet2001}
A.~Doucet, N.~de~Freitas, and N.~Gordon, {\em {Sequential Monte Carlo methods
  in practice}}.
\newblock Springer Science \& Business Media, 2001.

\bibitem{Godsill2004}
S.~J. Godsill, A.~Doucet, and M.~West, ``{Monte Carlo Smoothing for Nonlinear
  Time Series},'' {\em Journal of the American Statistical Association},
  vol.~99, pp.~156--168, mar 2004.

\bibitem{douc2011}
R.~Douc, A.~Garivier, E.~Moulines, and J.~Olsson, ``{Sequential Monte Carlo
  smoothing for general state space hidden Markov models},'' {\em Annals of
  Applied Probability}, vol.~21, no.~6, pp.~2109--2145, 2011.

\bibitem{wills2013}
A.~Wills, T.~B. Sch{\"{o}}n, L.~Ljung, and B.~Ninness, ``{Identification of
  Hammerstein–Wiener models},'' {\em Automatica}, vol.~49, no.~1, pp.~70--81,
  2013.

\end{thebibliography}

\end{document}